\documentclass[10pt, dvipsnames]{wlscirep}
\usepackage[utf8]{inputenc}
\usepackage[T1]{fontenc}
\usepackage{lineno}

\title{Emergence of universality in the transmission dynamics of COVID-19}
\author[a,b,$\dag$,$\ast$]{Ayan Paul}
\author[c,$\dag$]{Jayanta Kumar Bhattacharjee} 
\author[c,$\dag$]{Akshay Pal}
\author[d,$\dag$]{Sagar Chakraborty}

\affil[a]{Deutsches Elektronen-Synchrotron DESY, Notkestr. 85, 22607 Hamburg, Germany}
\affil[b]{Institut f\"ur Physik, Humboldt-Universit\"at zu Berlin, D-12489 Berlin, Germany}
\affil[c]{Indian Association for the Cultivation of Science, Jadavpur, Kolkata 700032, India}
\affil[d]{Department of Physics, Indian Institute of Technology Kanpur, Uttar Pradesh 208016, India}
\affil[$\ast$]{Corresponding author}
\affil[$\dag$]{e-mail:
\href{mailto:ayan.paul@desy.de}{ayan.paul@desy.de},
\href{mailto:jayanta.bhattacharjee@iacs.res.in}{jayanta.bhattacharjee@iacs.res.in},
\href{mailto:ug18ap@iacs.res.in}{ug18ap@iacs.res.in},
\href{mailto:sagarc@iitk.ac.in}{sagarc@iitk.ac.in}
}

\keywords{SARS-CoV-2, Transmission Dynamics, Universality Classes, Data Collapse}

\begin{abstract}
The complexities involved in modelling the transmission dynamics of COVID-19 has been a roadblock in achieving predictability in the spread and containment of the disease. In addition to understanding the modes of transmission, the effectiveness of the mitigation methods also needs to be built into any effective model for making such predictions. We show that such complexities can be circumvented by appealing to scaling principles which lead to the emergence of universality in the transmission dynamics of the disease. The ensuing data collapse renders the transmission dynamics largely independent of geopolitical variations, the effectiveness of various mitigation strategies, population demographics, etc. We propose a simple two-parameter model---the Blue Sky model---and show that one class of transmission dynamics can be explained by a solution that lives at the edge of a blue sky bifurcation. In addition, the data collapse leads to an enhanced degree of predictability in the disease spread for several geographical scales which can also be realized in a model-independent manner as we show using a deep neural network. The methodology adopted in this work can potentially be applied to the transmission of other infectious diseases and new universality classes may be found. The predictability in transmission dynamics and the simplicity of our methodology can help in building policies for exit strategies and mitigation methods during a pandemic. 
\end{abstract}

\begin{document}
\flushbottom
\maketitle

\thispagestyle{fancy}
\rhead{DESY 20-234, HU-EP-20/42}

\section*{Introduction}

The spread of SARS-CoV-2 has left significant instabilities in the socioeconomic fabric of the society. While the spreading dynamics of the disease is not novel~\cite{doi:10.1056/NEJMoa2001316}, the instabilities it has caused has made various parts of the society, and notably, various governance, respond to containing its spread in very different manners~\cite{Dehningeabb9789,10.1093/jtm/taaa020,SOHRABI202071,zhang2020covid,Fisher2020}. The determination of the optimal strategy has been quite a challenge and highly dependent on the socio-economic condition of the country or region~\cite{Fraser6146,Flaxman2020,2020arXiv201215230P,Kuhn2020.12.18.20248509}. A lot of effort has been spent trying to bring some predictability in the spread of the pandemic and even a few weeks of foresight can not only save an economy from being jettisoned but also save a considerable number of lives that need not be lost. Moreover, the experience of the past few months indicate that controlling the resurgence of the disease is a formidable task.

Various kinds of models have been used to describe the spread of COVID-19 with varying degree of success. These include several variants of the SIR (Susceptible, Infected , Recovered) model like SEIR~\cite{Zhou2020,Brett25897,Gatto10484, Bertozzi16732}---E for exposed, SEIRD---D for deceased, and Stochastic SIR~\cite{Lopez2020, Bertozzi16732}. There are more realistic delay differential equation models~\cite{Shayak2020.10.30.20223305,DellAnna2020,VYASARAYANI2020132701,Young2019,Kreck2021.01.18.21250012}, renormalization group equations~\cite{Cacciapaglia2020,10.3389/fphy.2020.00144,Cacciapaglia:2021vvu} and the highly computation intensive (and arguably the most detailed) agent-based network models~\cite{Wong29416,Chang2020,https://doi.org/10.1111/jep.13459,TATAPUDI2020100036,Rockett2020,Aleta2020,Hinch2020.09.16.20195925, Thurner22684,Thomas24180}. There are even purely data-driven models~\cite{Merow27456,Perkins22597,mkv}. In all these approaches, sets of parameters have to be tuned precisely to analyse a particular country or region to obtain predictions, since every region seems to have its own special time evolution for disease propagation, making this a Herculean task. Although all models' assumptions cannot strictly describe every detail of any real-world system, it is believed that such systems may possess certain universal behaviours independent of these details~\cite{Siegenfeld16092}. 

We propose to establish that there is a universality in the spreading of this disease which can be seen by subjecting the available data of all the countries or regions to a scaling analysis reminiscent of the data collapse in the study of critical phenomena~\cite{RevModPhys.49.435,RevModPhys.71.S358} in the sixties and seventies of the previous century. Data collapse leading to universality has been observed in a large variety of systems and we provide a short overview of the literature in the Appendix. We emphasize that the procedure we follow is model-agnostic and does not require the assumption of a model for the transmission and mitigation dynamics. In addition, this method lies open to model-based interpretations as we later point out. Interestingly, we find that the universal curve is well described by a simple, yet non-trivial, mean-field model where the carrying capacity of the infected population depends on the population size. The model---that we call Blue Sky model (BSM)---undergoes a blue sky bifurcation~\cite{AS1988} (also called the saddle-node bifurcation or the fold bifurcation) which is a ubiquitous, structurally stable bifurcation found in variety of systems, such as power systems~\cite{RAO20171}, viscous profiles~\cite{ACHLEITNER20121703}, neural networks~\cite{cite-key}, excited molecules~\cite{doi:10.1021/jp0131065}, plasmas~\cite{doi:10.1063/1.3435269}, and the Ginzburg--Landau systems~\cite{PhysRevE.67.015601}. At the edge of the blue sky bifurcation point, appears a universal data collapse. Furthermore, given that the scaling and data collapse have stood well the test of time for a very diverse set of systems and is grounded in well understood physical principles, we are proposing a robust method for making long-term predictions that will help in evaluating disease mitigation efficacies and planning exit strategies for the future. We hope to convince the readers that we have established a method which does not require parameter tuning or model selection to gain a handle over the prediction of disease spread.

The questions that we try to address in our work are as follows:
\begin{itemize}
    \item Even though the pattern of disease spread is diverse in different geographical regions due to several important factors that include local socioeconomic conditions~\cite{Paul_2021}, population demographics and dynamics~\cite{Wilder25904}, mobility patterns~\cite{Nouvellet2021} and factors like climate, pollution, etc.~\cite{ROSARIO2020113587,DIAO2021100203,RAHIMI2021110692,cite-key-Wang,cite-key-Yechezkel,COCCIA2021145801,COCCIA2021437}, is it possible to rescale the data so that the universality in the transmission dynamics can be extracted?
    \item Can this universality lead to greater predictive power for certain phases of the disease spread and allow the explanation of the transmission dynamics in terms of a simple universal model leading to a vast reduction of the number of parameters necessary for predicting the spread of the disease?
\end{itemize}

The primary contributions of our work are:
\begin{itemize}
\itemsep0em
    \item We show that there exists a universality in the spread of the disease once proper scaling is applied to the data. 
    \item Two universality classes emerge that give us a window into understanding how well a country or region performed in mitigating the disease. These universality classes are independent of geographical scales.
    \item We propose the BSM---an extension of the logistic growth model---which explains the existence of these two classes, the first, through a solution that lives at the edge of blue sky bifurcation and a second that maps the BSM onto the logistic growth model. To our knowledge, this model has not been discussed in the literature before.
    \item We show how predictability can be realized from this analysis in a model-agnostic manner relying only on the inference drawn from the first phase of the disease spread. For this, we construct a deep neural network (DNN) based predictive model. We also show that an equally good prediction can be made using the BSM. We give a few examples from the ongoing phase of the pandemic in several countries and states.
    \item Our work focuses more on explaining the progression of the spread past the peak on the daily case rates for infection since the initial growth is well explained by the logistic model and some of its extensions that we discuss in the Appendix.
\end{itemize}

There have been a few recent attempts to explore universality in the spreading of COVID-19. In Ref.~\cite{doi:10.1098/rspa.2020.0689} the focus is on the early exponential growth of the disease and the analogy with the growth of the phase separation following a quench from the one-phase to the two-phase region on either side of the critical point of a fluid. In Ref.~\cite{10.1093/ptep/ptaa148}, the stress is on the asymmetric dependence of the daily infection rate on time. The claim made in this reference is that the tail of the distribution can be described by the double exponential Gompertz function. This is to be contrasted with the power-law tail which can be inferred from Ref.~\cite{Sharma2021}. Other works have focused on extracting universal quantities that quantify the dynamics of disease spread appealing to  renormalization group techniques~\cite{Cacciapaglia2020,10.3389/fphy.2020.00144,Cacciapaglia:2021vvu}. In the following sections we will describe in detail our method of scaling the data that leads to the emergence of universality and the consequent predictability gained.

 \begin{figure*}[ht!]
	\centering
	\includegraphics[width=\textwidth]{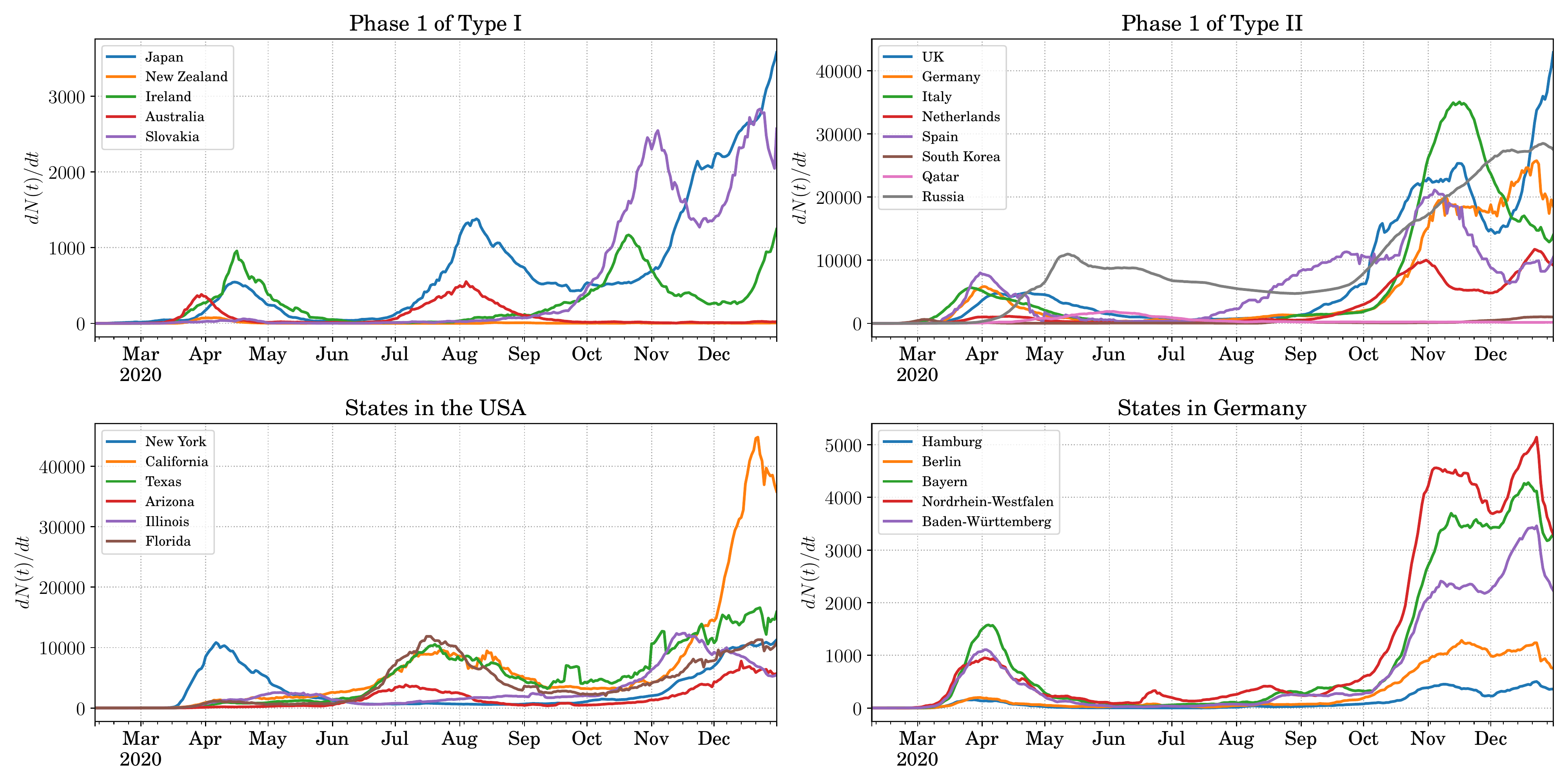}
	\caption{\it \textbf{The variation in the spreading of COVID-19 in various countries and some states of the USA and Germany.} {Top left:} the countries for which the first phase of spreading falls into the category defined as Type I. {Top right:} the countries for which the first phase of spreading falls into the category defined as Type II. {Bottom left:} some states in the USA that show very different spreading patterns for the disease. {Bottom right:} some states in Germany that show very similar pattern in the spreading of COVID-19.}
	\label{fig:countrydata}
\end{figure*}
\section*{Emergence of Universality in Transmission Dynamics}
\label{sec:transuniversality}

To bring about some semblance of order in what looks like a disparate set of data from several countries, 
we first focus entirely on phase-1 of the pandemic. We denote the total number of cases (individuals who have tested positive for the COVID-19) at time $t$ as $N(t)$ with $t$ being measured in days. The derivative $dN(t)/dt$ is the number of new cases detected on a given day. Phase-1 refers to the situation, for a given country or region, where the $dN(t)/dt$ increased from zero or a small value, went through a maximum and then decayed to almost zero (Type I transmission) or to a small value (Type II transmission). It stayed near zero or near the small value for a few weeks before starting to rise again at the onset of the next phase. We need to differentiate quantitatively between the phrase ``near zero for a few weeks'' and ``a small value for a few weeks''. We do this by inspecting the rise in the total number of cases, $N(t)$, over four weeks past phase-1. If the change is more than 4\% of $N(t)$, we mark the transmission as being of Type II, otherwise it is Type I.

Taking a more careful look at \autoref{fig:countrydata}, we note that the total number of cases and the total duration of phase-1 are very different for each country. This is to be expected as these countries have very different population densities and different patterns of inhomogeneities in the density of population. Moreover, the policies of the government and the acceptance of these by the governed affect the implementation of precautionary measures~\cite{Flaxman2020,2020arXiv201215230P,Kuhn2020.12.18.20248509,Mitze32293}. The policies and exit strategies towards the end of a phase in any country also determines whether the spread of the disease lingers on, albeit, at a much reduced rate, or is completely routed. The latter essentially divides the world into two classes in our formalism. However, looking at bare case numbers, one is bound to conclude that each country or region is a unique case study requiring a detailed modelling of transmission dynamics and incorporating the effects of mitigation and containment measures, a task that seems formidable right from the onset.

The top left and right panels of \autoref{fig:countrydata} show some examples of countries that went through a phase-1 transmission of Type I and Type II respectively. The disparity in the spreading patterns is apparent. In the lower panels of \autoref{fig:countrydata} we focus on the states of the USA and Germany. In the USA we see that the spreading pattern in the different states was somewhat decorrelated. This is where our formalism is particularly useful. The physical principles that we advocate being independent of geographical scales can be applied to countries, states or even counties and districts given they have seen a significant spread of the disease within the region. On the other hand, the spread of COVID-19 in Germany in the various states has been quite correlated. What is important to note here is not only are the scales of the disease spread different, but also the shapes of the distribution over time vary considerably, hinting at various timescales characterizing each region. Hence a key part of our work will be to bring about a data collapse in both these dimensions which cannot be achieved by simple rescaling the numbers alone.

Now, let us see how universality can emerge from a scaling of the COVID-19 confirmed case rate data. We introduce a number $N_{\rm max}$ which is the total number of infected people at the end of phase-1. For Type I transmission the value of $N_{\rm max}$ is easy to infer. It is the the total number of cases at which the $N(t)$ curve flattens out after the close of the phase. For Type II transmission, where $N(t)$ is never completely flat, we determine the close of phase-1 following the procedure described above. A perfect example is the case of Germany. For the period May 31$^{st}$, 2020 to July 17$^{th}$, 2020, the value of $dN(t)/dt$  hovered around 400 which is more than an order of magnitude smaller than the peak value of 6000. In this case $N_{\rm max}$ corresponds to the number of cases at the end of this period of stagnancy in the growth rate, i.e., the number of cases on July 17$^{th}$, 2020. Of course, there are some degree of variability associated with the choice of the end-date for a phase. However, an estimation error of a few days does not change the onset of universality. Therefore, we choose the end-date for phase-1 as the first day when the $N(t)$ curve flattens out for Type I transmission and the last day before the onset of the next phase for Type II transmission.

Our conjecture is that universality emerges from the scaling given by
\begin{eqnarray}
\tilde t = \frac{t}{t_{1/2}}, \;\;{\rm and}\;\; \tilde N (\tilde t)=\frac{N(\tilde t)}{N_{\rm max}}.
\label{eq:r}
\end{eqnarray}
The ratio $\tilde N (\tilde t)$ is relatively insensitive to the testing policies, the inhomogeneity in the population density, the size of the country or region, the details of the mitigation methods advocated, etc. The rescaling of the time by an appropriate time scale for each country or region, irons out the nuances brought about by the government policies like the duration of lockdown, exit strategies, etc., at the lowest order of approximation. Therefore, for each country or region a natural time scale to choose is $t_{1/2}$  which is the time at which $N(t) = N_{\rm max}/2$.

The choice of the characteristic timescale is subjective. For instance, choosing $t=t_{1/n}$ corresponding to the time when $N(t)=N_{\rm max}/n$, with $n$ being a positive definite number would not affect the nature of this analysis. However, choosing $n=2$ allows for an easy interpretation of the characteristic time scale and, as we see later, places the peak of $d\tilde N (\tilde t)/ d\tilde t$ at $t=t_{1/2}$ for a simple logistic growth. According to our conjecture, $\tilde N (\tilde t)$ plotted against $\tilde t$ should show a data collapse opening up avenues for simple predictive modelling and generalization beyond socioeconomic and geopolitical variations.

\begin{figure*}[t!]
	\centering
	\includegraphics[width=1\textwidth]{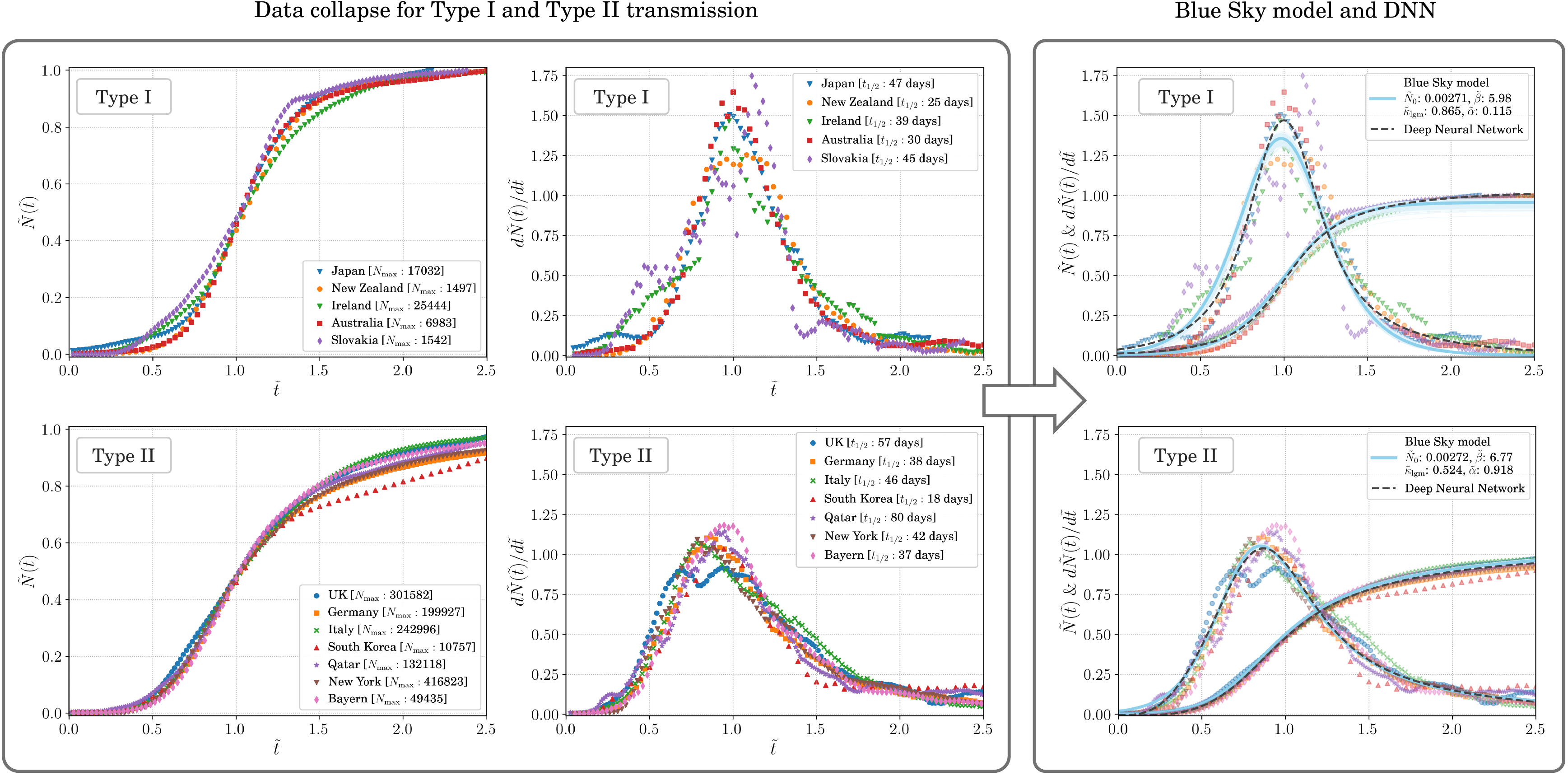}
	\caption{\it \textbf{The emergence of universality in the transmission dynamics of COVID-19}. {Left panels:} The rescaled total number of confirmed cases, $\tilde N(\tilde t) = N(t)/N_{\rm max}$, and the its derivative, $d\tilde N(\tilde t)/d\tilde t$, with respect to the scaled time $\tilde t = t/t_{1/2}$ for Type I transmission where the disease spread was almost completely routed after phase-1 and Type II transmission where the disease spread was not completely routed after phase-1. The data collapse is evident and independent of the country or state being considered showing the emergence of two universality classes in the transmission dynamics of COVID-19 independent of social and geopolitical variations. {Right panels:} The fit to the rescaled data using the Blue Sky model (solid blue line) and the model agnostic fit (dashed black line) with a deep neural network. Note that $\tilde\alpha=0.918$ in the rightmost bottom panel is practically the value expected at the edge of the blue sky bifurcation at $1/4\tilde\kappa_{\rm lgm}^2=1/(4\times0.524^2)=0.910$. The good agreement between the two approached can be clearly seen.}
	\label{fig:universal}
\end{figure*}

\subsection*{Data-collapse in transmission dynamics of COVID-19}
\label{sec:covid19}

In the light of the preceding discussion we present the $\tilde N(\tilde t)$ vs. $\tilde t $  plots for several countries and states in left-most plots of \autoref{fig:universal}. The data collapse is clearly evident. We see that independent of variabilities in transmission dynamics the universality in the rescaled data dictates that there are two distinct classes. The plots in the middle of \autoref{fig:universal} show the functional dependence of $d\tilde N(\tilde t)/d\tilde t$ on $\tilde t$. For the Type I transmission one clearly sees that tail of the distribution at higher $\tilde t$ dips to very low values. This implies that after phase-1 ended, the daily case rate actually went down to a negligible number, which, originally, was our basis for identifying Type I transmission. On the other hand, for the same curve in the lower middle plot of \autoref{fig:universal}, the tail of the distribution at higher $\tilde t$ never really goes to as low values as that for Type I transmission. This is because in a transmission of Type II the countries and regions never saw a complete routing of the spread of the disease. To show the independence of geographical scales of our formalism, we include New York, USA and Bayern, Germany as two states tracking them over phase-1 of the spread of COVID-19. We see that both the states show transmission of Type II.

It is important to point out here that the data collapse that we see effectively filters out the variances in the dynamics of the disease spread and the policies for disease mitigation. It also effectively filters out how quickly the response to the pandemic was orchestrated (a matter of government policy), how drastic these measures were and all other geopolitical demographics that make the disease propagation dynamics in each country or state look very different in \autoref{fig:countrydata}. The key difference between the two classes reduces to the effectiveness of the countries in fully containing the disease itself, each using their own set of mitigation methods and exit strategies. Indeed, this data collapse removes all necessity of fine-tuning a large number of parameters to examine each country or state separately and shows that there is a common pattern in the spread of COVID-19.

\subsection*{Interpretation of $\mathbf{t_{1/2}}$}
\label{sec:t12}
The parameter $t_{1/2}$ gives us a distinct insight into how well the spread of the disease was mitigated in a particular phase. A shorter $t_{1/2}$ implies a more effective mitigation strategy. For example, from \autoref{fig:universal} we see that the $t_{1/2}$ for South Korea is 18 days, which means that the spread of the disease was stopped very rapidly as compared with UK for which $t_{1/2}$ is 57 days. However, despite being effective in slowing down the spread rapidly, South Korea falls into Type II since the spread was not completely stopped but lingered on. On the other hand, Ireland, even being of type I, has a relatively longer $t_{1/2}$ of 39 days showing that while the spread of the disease was almost completely routed it took a longer period of time to do so hinting as possibilities of drafting better mitigation strategies.

\subsection*{Blue Sky model}
\label{sec:BlueSkyModel}
Among the four broad classes of models (with possibility of overlap among them) of infectious disease spreading---viz, data-driven phenomenological models~\cite{LEGA201619, cite-key786}, compartmental or lumped parameter models~\cite{Brauer2008}, agent-based models~\cite{2020Natur.580..316A}, and stochastic differential equation models~\cite{Allen2008}---while the former two are mathematically simpler, the latter two are more detailed in their approach. The first two types of models are to the last two types, what thermodynamics is to statistical mechanics. The most simple lumped parameter model is undoubtedly the well-known one dimensional autonomous logistic growth model~\cite{Verhulst}:
\begin{eqnarray}
\frac{dN(t)}{dt}=\beta N(t)\left(1-\frac{N(t)}{\kappa}\right),
\label{eq:lgm}
\end{eqnarray}
from which we shall begin. Here, $\beta>0$ is the Malthusian growth rate in the presence of unlimited resource, i.e., $\kappa\rightarrow\infty$. Most commonly, the resource is assumed to be unaffected by the population size, $N(t)$, and hence, the population grows to settle onto the value $N(\infty)=\kappa$, which is the carrying capacity. We discuss some variants of the the logistic growth model and their fit to the data in the Appendix. 

The system of our interest is a population where $N(t)$ represents the number of COVID-19 infected individuals that can potentially infect healthy individuals so that the infected population grows at a rate $\beta$. All the depletable conditions that are favourable for the infected population to grow is quantified through the positive parameter, $\kappa$. One could argue that rather than relying completely on the carrying capacity parameter, one could make this model more elaborate by making it higher dimensional to mathematically incorporate the interaction among infected, susceptible, exposed, asymptomatic, quarantined, recovered, immune, and deceased individuals. However, the fact remains that all such higher dimensional models, despite their share of success, bring in more parameters while the only available data is still mostly that of the infected and deceased individuals. Thus, in this paper, we are motivated to take the route of simple two-parameter one dimensional model complimented with the information extracted from the real data.

It is straightforward to realize that the carrying capacity can easily depend on $N(t)$: e.g., the government may practically be incapable of quarantining infected individuals as their numbers grow, thereby effectively increasing the carrying capacity because the infected individuals would interact with more and more healthy individuals leading to terms which are higher order in $N(t)$. Therefore, if we were to model the realistic scenarios, we should let $\kappa$ depend on $N(t)$ in a nontrivial way. In this context, we find it useful to define an order parameter, $\rho$, as
\begin{eqnarray}
{\rho}\equiv \frac{1}{\kappa^2}\left(\frac{d\kappa}{d{N}}\right).
\label{eq:OP}
\end{eqnarray}
Note that lower the value of the order parameter, more is the control on the change of the carrying capacity. The case of the order parameter being zero corresponds to the traditional logistic equation governed system where the carrying capacity is independent of $N(t)$. We would like to mention here that past considerations~\cite{cite-key1,SHEN2020582,Shayak2020.10.30.20223305,cite-key3,AVIVSHARON2020502,9138049,MALAVIKA202126,AllahHuAkbar,PELINOVSKY2020110241,WANG2020110058} of the logistic growth related compartmental models, as applied to the modelling of COVID-19, do not explicitly incorporate the natural notion of the state-dependent carrying capacity.

In order to account for the aforementioned variation of the carrying capacity with the change in the infected population size, we propose a simple, yet non-trivial, functional form for the carrying capacity:
\begin{eqnarray}
{\kappa}=\kappa_{\rm bsm}\equiv\frac{{\kappa_{\rm lgm}}}{1-{\alpha}{\kappa_{\rm lgm}}{ N(t)}}\,,
\label{eq:BSM-CC}
\end{eqnarray}
where ${\kappa_{\rm lgm}}$ is the constant carrying capacity corresponding to the traditional logistic growth model and $\alpha$ is a non-negative parameter. In \autoref{eq:lgm} with $\kappa_{\rm bsm}$ as the carrying capacity, we define the {Blue Sky model (BSM)} that we use to explain our findings in this paper. In the BSM, we note $\rho=\alpha$ which renders a significant physical meaning to the (otherwise ad hoc) parameter $\alpha$. Following the transformations, $\tilde{N(t)}=N(t)/N_{\rm max}$ and $\tilde{t}=t/t_{1/2}$, adopted in the previous discourse, we define $\tilde{\beta}\equiv t_{1/2}\beta$, $\tilde{\kappa}_{\rm lgm}\equiv \kappa_{\rm lgm}/N_{\rm max}$, and $\tilde{\alpha}\equiv N_{\rm max}^2\alpha$. Henceforth, we shall consider only the rescaled BSM where these rescaled parameters are used. 

In the BSM, the blue sky bifurcation occurs at the nonzero fixed point at $\tilde\alpha=\tilde\alpha_{\rm edge}\equiv1/4\tilde\kappa_{\rm lgm}^2$---\emph{the edge of blue sky}. Beyond the edge of blue sky, i.e., for $\tilde\alpha>1/4\tilde\kappa_{\rm lgm}^2$, the two nonzero fixed points vanish and $\tilde N(\tilde t)=0$ is the sole fixed point in the system which is unstable. In other words, the population size of the infected individual grows without any bound, which is a physical impossibility. This sets an important limit that $\tilde\alpha$ should respect for the model to be physical: one must impose the condition, $\tilde\alpha\in[0,1/4\tilde\kappa_{\rm lgm}^2]$, so that a stable attracting nonzero state of the infected population, viz., $\tilde N(\tilde t)=(1-\sqrt{1-4\tilde\alpha\tilde\kappa_{\rm lgm}^2})/(2\tilde\alpha\tilde\kappa_{\rm lgm})$ exists. The third fixed point, $\tilde N(\tilde t)=(1+\sqrt{1-4\tilde\alpha\tilde\kappa_{\rm lgm}^2})/(2\tilde\alpha\tilde\kappa_{\rm lgm})$, is unstable and not of any practical interest to us.

As seen in the rightmost column of \autoref{fig:universal}, the BSM remarkably segregates the COVID-19 transmission dynamics in countries and states into two universality classes---Type I and Type II---characterized by two distinct and well separated values of the (rescaled) order parameter, $\rho$---one close to the logistic growth value ($\tilde\alpha\approx0$) and the other close to the edge of blue sky ($\tilde\alpha\approx1/4\tilde\kappa_{\rm lgm}^2$) respectively. The data collapse onto the respective curves is strongly reminiscent of the similar phenomenon observed in the critical phenomena and their universality across a plethora of physical systems~\cite{RevModPhys.49.435,RevModPhys.71.S358}. Thus, we speculate that it may be a potential hint of a hidden renormalization scheme present in the microscopic details of the corresponding agent-based models.

\begin{figure*}[t!]
    \centering
    \includegraphics[width=0.32\textwidth]{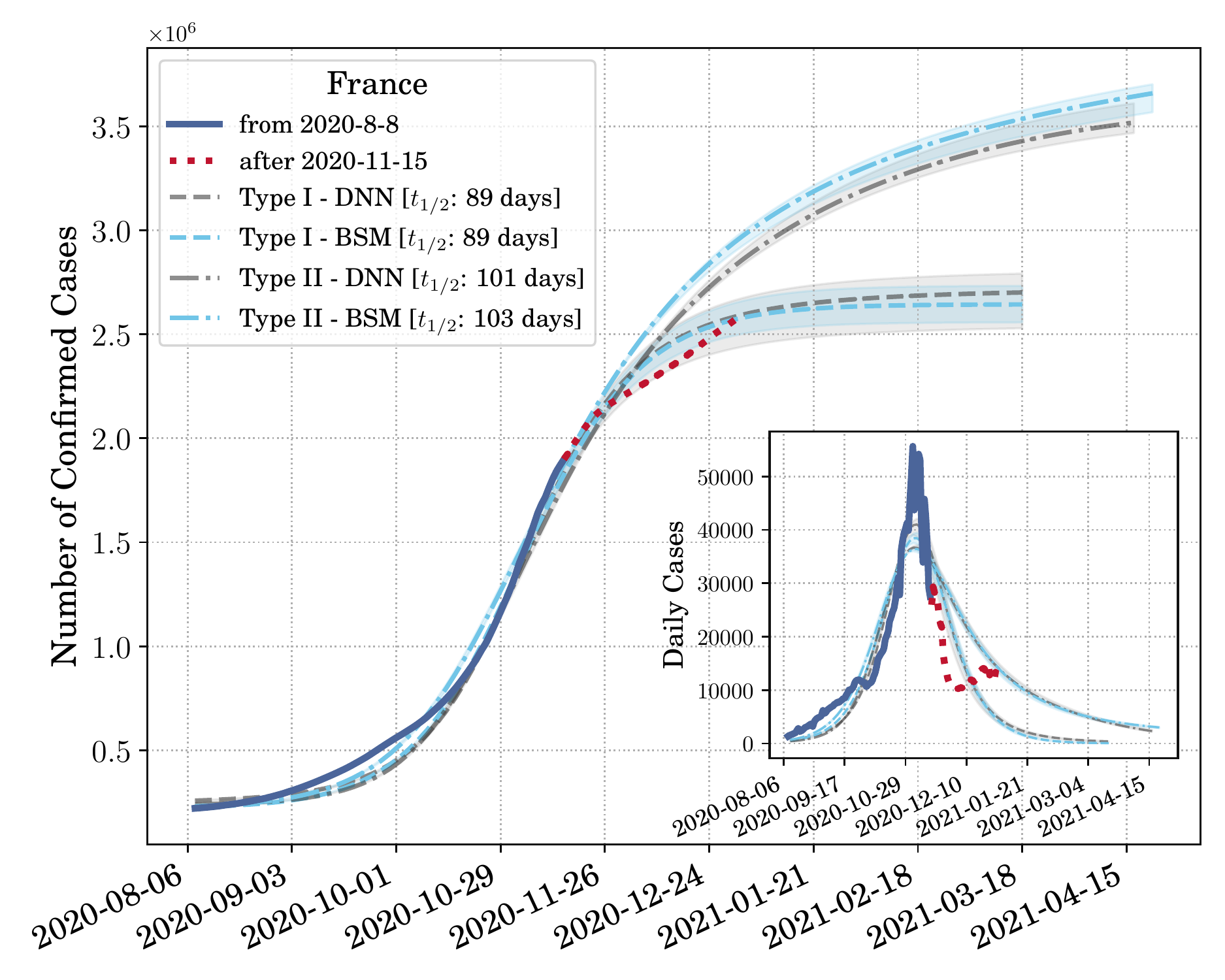}
    \includegraphics[width=0.32\textwidth]{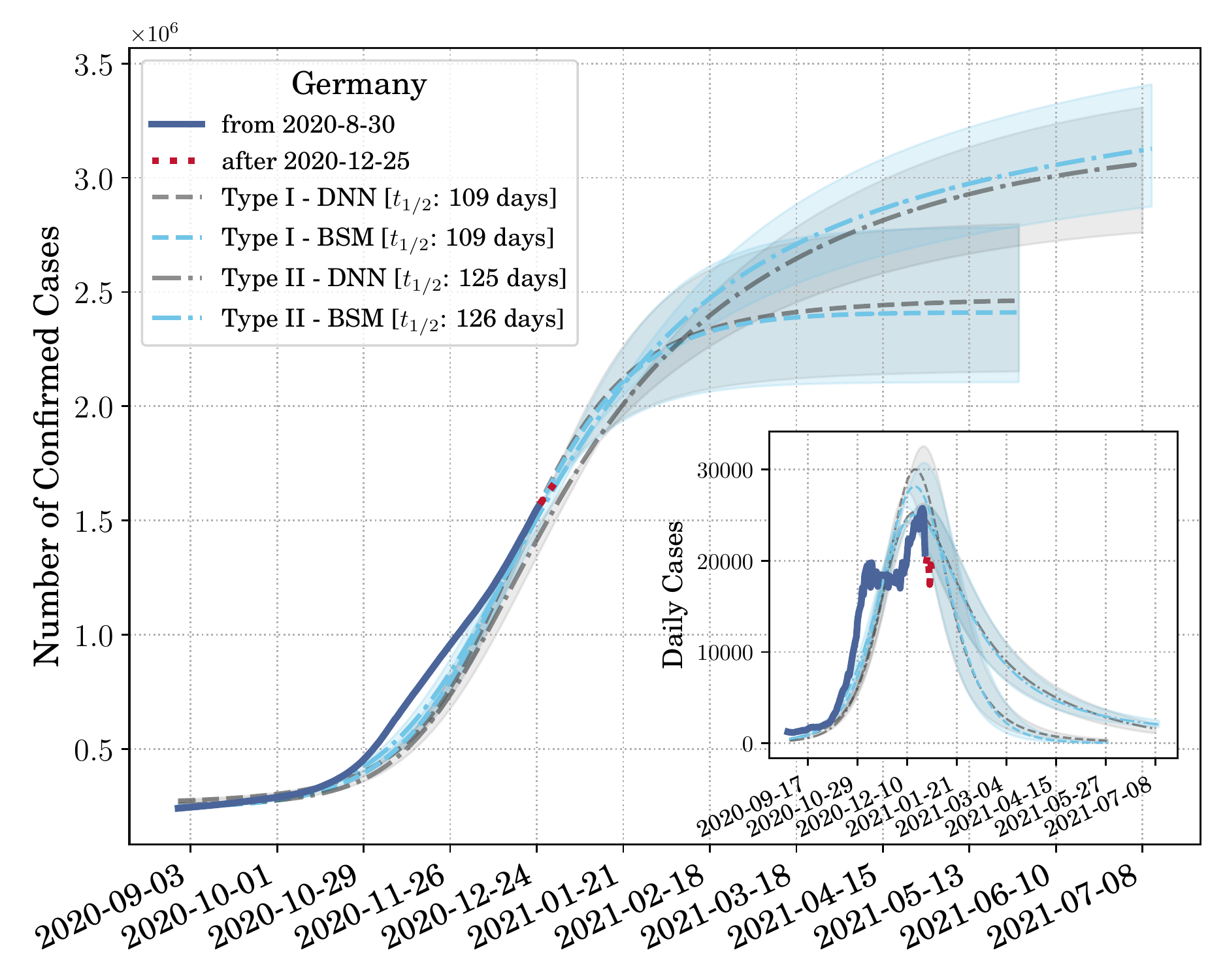}
    \includegraphics[width=0.32\textwidth]{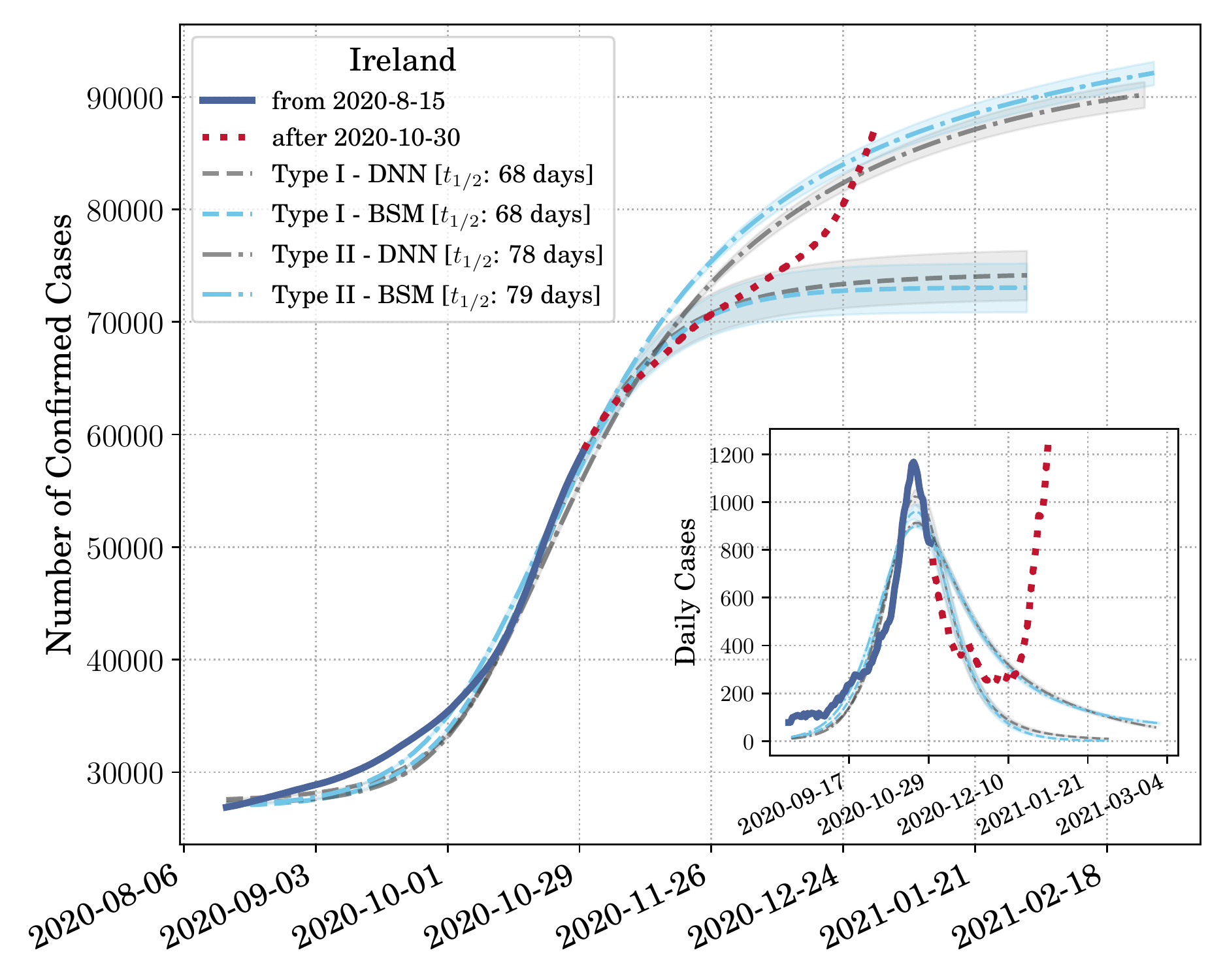}
    \includegraphics[width=0.32\textwidth]{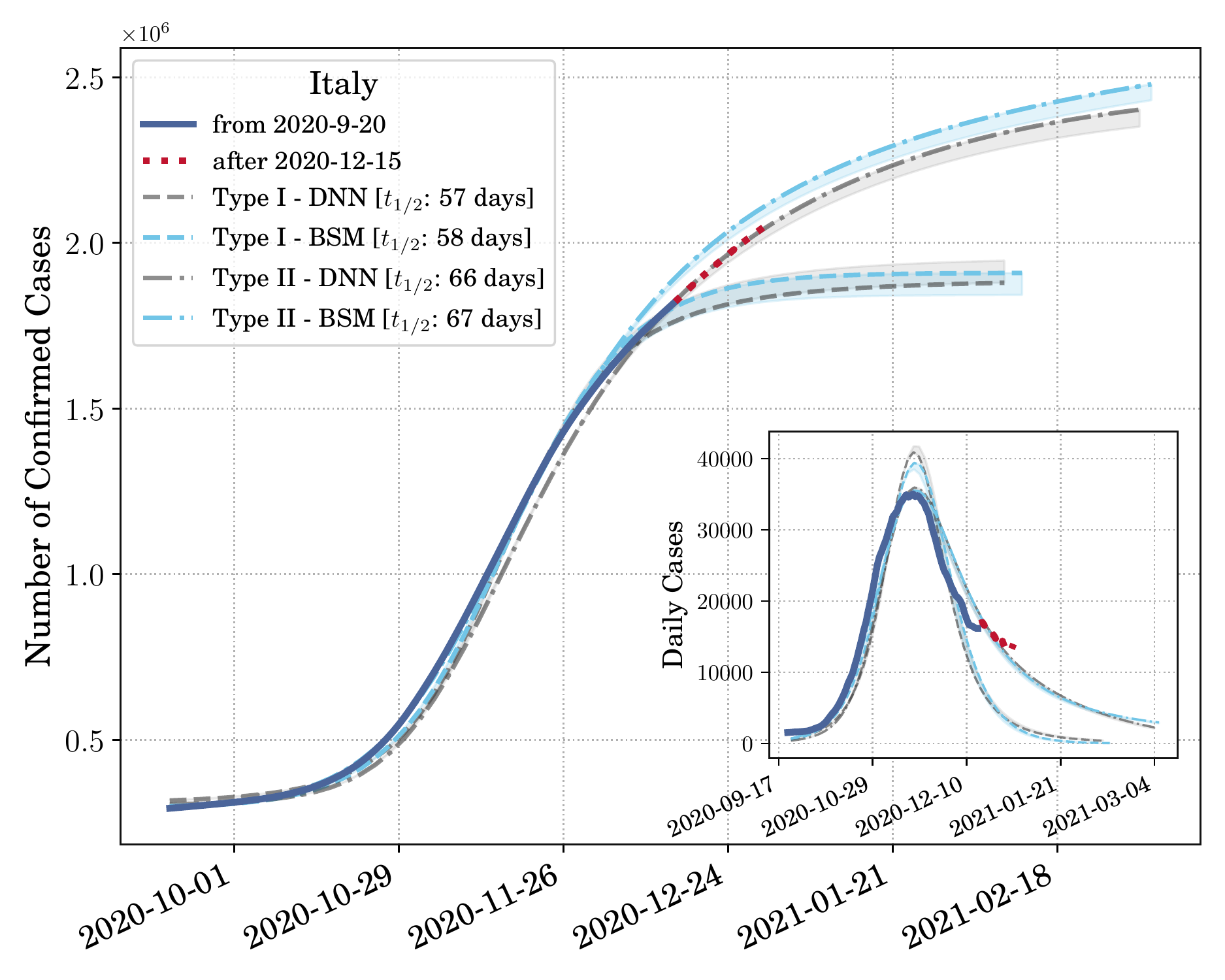}
    \includegraphics[width=0.32\textwidth]{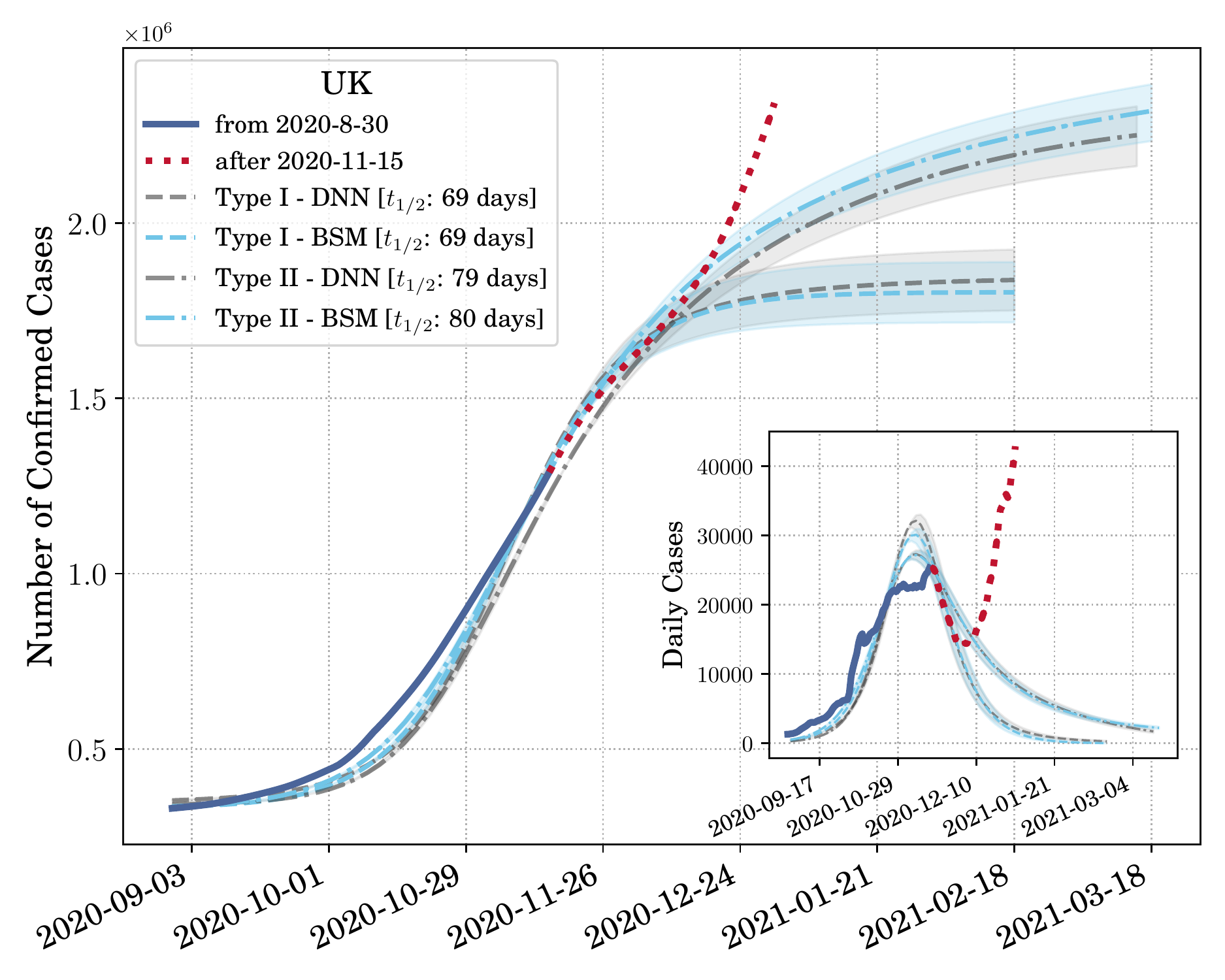}
    \includegraphics[width=0.32\textwidth]{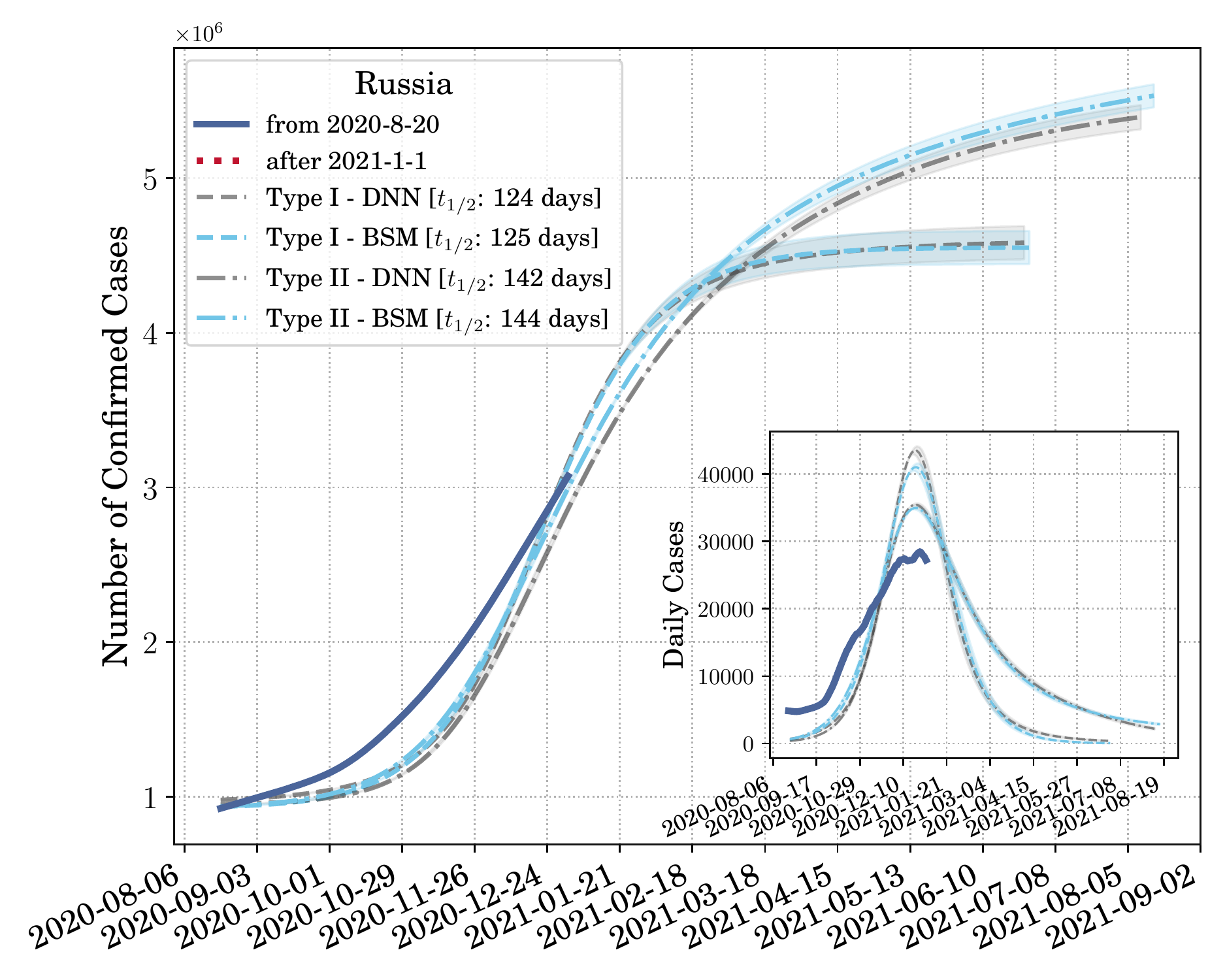}\\
    \includegraphics[width=0.32\textwidth]{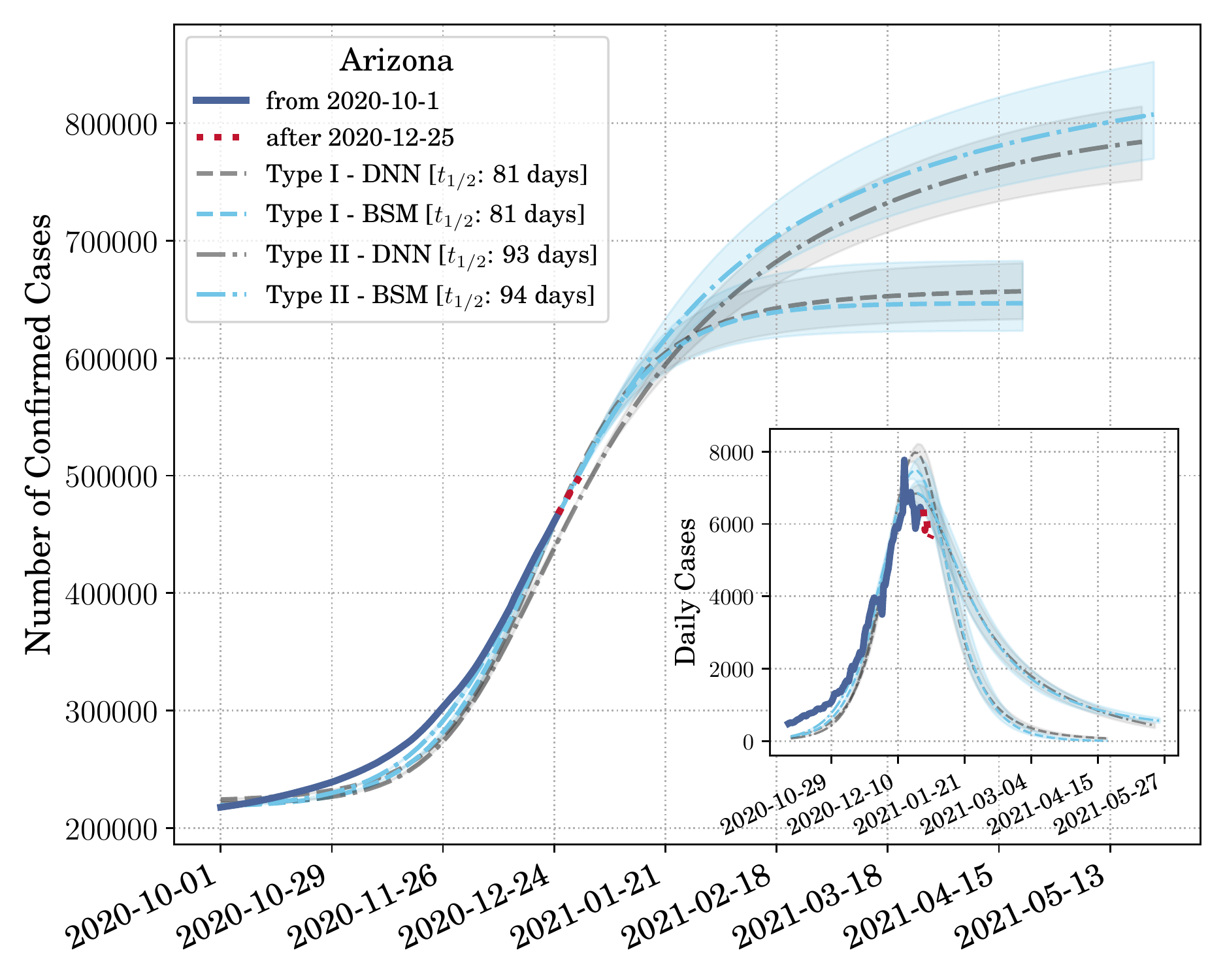}
    \includegraphics[width=0.32\textwidth]{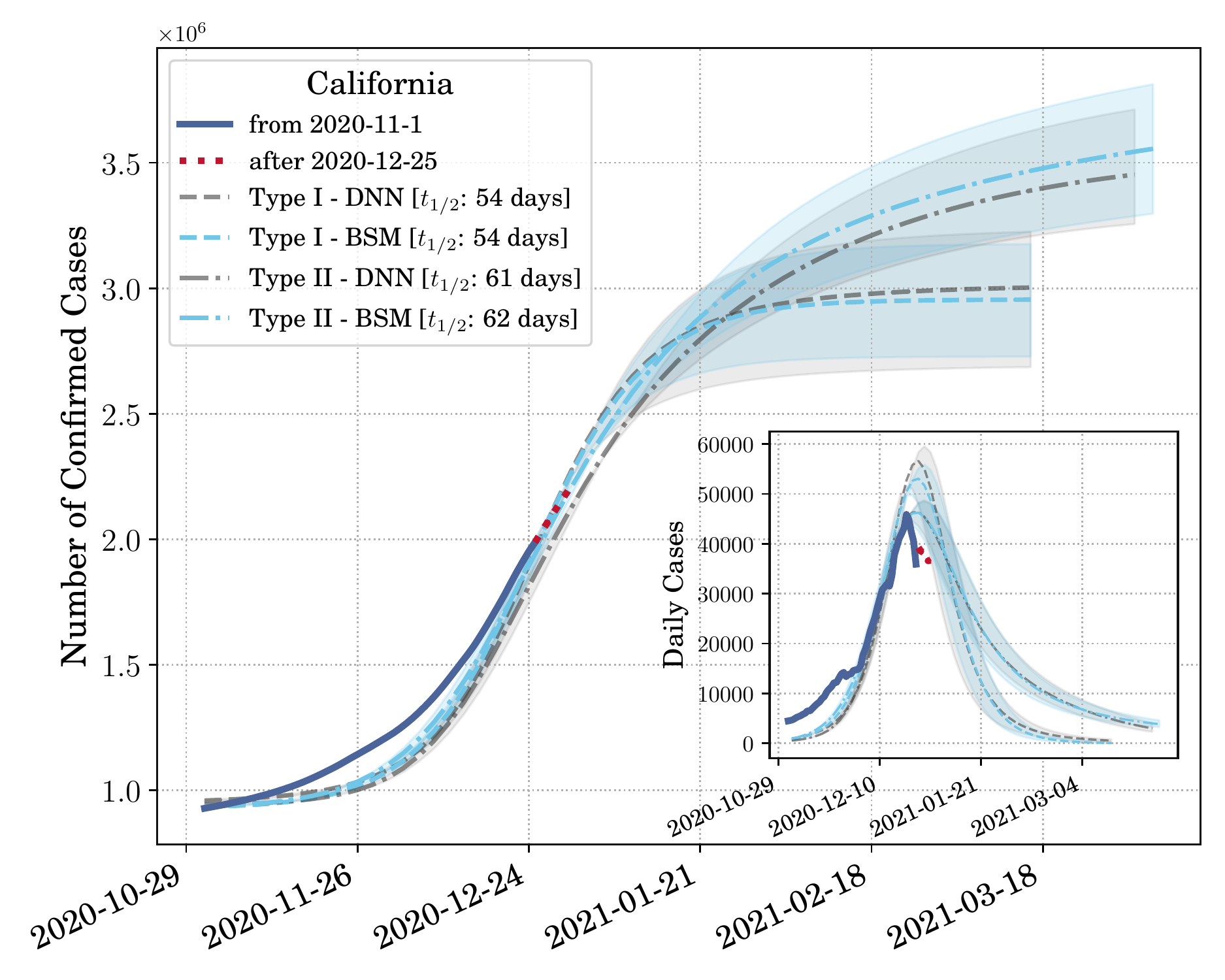}
    \includegraphics[width=0.32\textwidth]{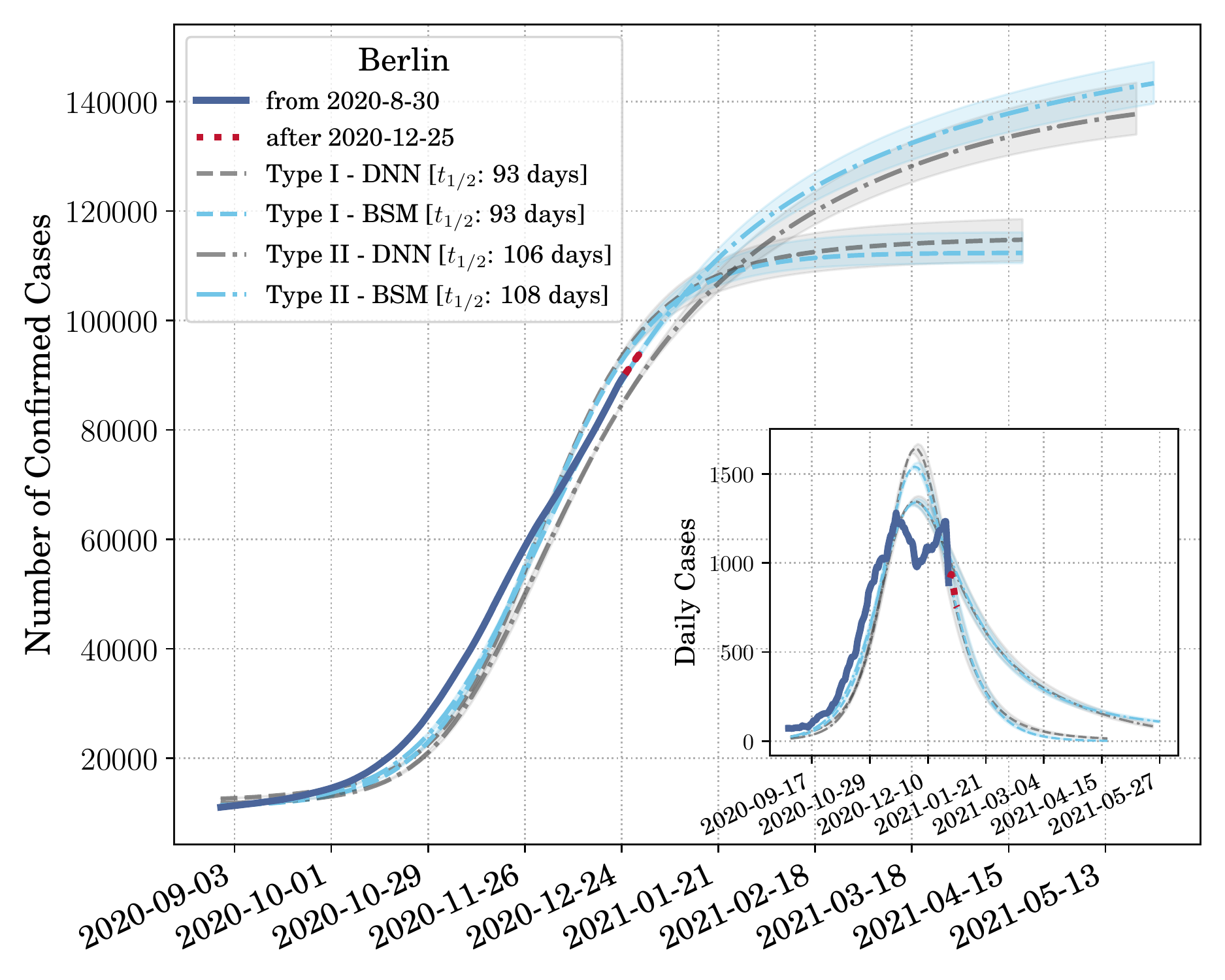}
    \caption{\it \textbf{A display of predictability from the emergence of universality classes.} We show how predictions can be made form scaling data and using knowledge gathered from phase-1 of the pandemic. Once the peak in the daily case rate is reached (or has been assumed to have reached), prediction for several weeks or months can be made. The solid dark blue lines in the insets show the data being used to find $t_{\rm peak}$, and from it, $t_{1/2}$ and $N_{\rm max}$. The dotted red lines in the insets are the test data which lie in the future with respect to the data used to determine the peak. The light blue lines and bands are predictions from the model-dependent fit using the Blue Sky model (BSM) to phase-1 data and the gray lines and bands are those from the model-agnostic fit using a deep neural network (DNN). The dashed lines are prediction for transmission of Type I and the dot-dashed lines are for transmission of Type II. All predictions are made till $\tilde t = 2.5$ or $t=2.5t_{1/2}$ starting from the beginning of the ongoing phase of disease spread. The top and middle rows show predictions for various countries while the bottom row shows predictions for two states in the USA and one in Germany.}
    \label{fig:pred}
\end{figure*}

\subsection*{Deep neural network}
\label{sec:DNN}

While the BSM provides a very good and intuitive explanation of the data, we feel that one should not be limited to exploring the universality in the data through specific models only. Hence, we perform a model-agnostic fit using a universal function generator in the form of a deep neural network (DNN). We use a three-layers deep, with each layer 16 nodes wide DNN for this purpose. The core of the architecture can be found in the Methods and Materials section and a detailed discussion can be found in the Appendix. The fit to the data using the DNN can be seen in the rightmost panel of \autoref{fig:universal}. The back dashed line corresponds to the DNN fit. We fit to the $\tilde N (\tilde t)$ distribution. We see that the fit of Type II transmission is almost identical to the BSM fit showing that the data-driven and model-dependent methods reach the same conclusion. For transmission of Type I the DNN fit is slightly different from the BSM fit and notably so at the center of the distribution where $d\tilde N (\tilde t)$ is shown to be higher from the DNN fit than from the BSM fit.

\section*{Predicting of COVID-19 spread from universality}
\label{sec:predictions}

The data collapse that we show leads us to the ability of making predictions in both a model-dependent and a model-agnostic way. While the BSM shows a clear explanation of the data, the advantage of a data collapse lies in the emergence of universality which allows for model-agnostic predictions from the simple understanding that the transmission dynamics described by the rescaled variables are the same in every phase of the the disease spread. Hence, we would like to emphasize both possibilities although they lead to similar results in predictions.

It should be noted that the only parameter that needs to be estimated for predictions to be made is $t_{1/2}$ for the ongoing phase for a country or region. Once that is determined, the parameters extracted from the fit to the phase-1 data can be used to make predictions for the ongoing phase. For both the approaches we use a common algorithm except for the functional form that we use to fit to the data. For the model-dependent approach we use the BSM and extract its parameters from data, whereas in the model-agnostic approach we use the trained DNN to predict the distribution of $\tilde N (\tilde t)$. 

The algorithm used to make the predictions is delineated in Materials and Methods section. To show the extent to which our prediction work, we assume that a country or state is just past the peak of $dN(t)/dt$. The data that we use for determining $t_{\rm peak}$ is represented by the solid dark blue line in the insets of \autoref{fig:pred}. The data past the end of this line, marked with a dotted red line is assumed to be in the future. We use the ``past'' data to predict for the ``future'' and then extend our prediction to $\tilde t = 2.5$. Since $t_{1/2}$ is unique for each country or state, each approach and each transmission types, the length of time for which predictions are made differ.

\subsection*{Model-dependent prediction}

The fitting of the BSM to data from phase-1 paves the way for making model-dependent predictions for future phases of the pandemic. This can be done by identifying $t_{1/2}$ and, from it, $N_{\rm max}$. This is only possible if, in a current phase of the pandemic, the daily case rate $dN(t)/dt$ has already reached a peak. This peak lies at time $t=\tilde t_{\rm peak} t_{1/2}$ where $\tilde t_{\rm peak}$ can be derived from the fit parameters and is given in \autoref{eq:BSMtmax}. By definition, $t=t_{1/2}$ is the time at which the $N=N_{\rm max}/2$ with $N(t)$ being counted from the beginning of the phase. Once these two parameters have been extracted, one can make two separate predictions for any geographical region: the first assuming a spread of Type I and the second assuming a spread of Type II. These predictions for Type I (dashed blue line) and Type II (dot-dashed blue line) are shown in \autoref{fig:pred} for several countries and states. The bands in blue represent the error in the prediction propagated from the error in the estimation of $\tilde t_{\rm peak}$.

\subsection*{Model-agnostic prediction}

We also show the predictions from the model-agnostic approach using a trained DNN in \autoref{fig:pred}. The logic that we follow to identify the peak of the $dN(t)/dt$ distribution is the same as for the model-dependent predictions. However, $t_{1/2}$ now varies between the two methods since $\tilde t_{\rm peak}$ is different for the model-dependent and the model-agnostic fits. The dashed gray lines represents transmission of Type I and the dot-dashed gray lines represents transmission of Type II. For Type I transmission, the BSM gives predictions similar to the DNN. However, for Type II the BSM mostly predicts somewhat larger number of cases. This is because $\tilde t_{\rm peak} = 0.85$ from the BSM fit of the phase-1 data is a bit lower than that from the DNN fit of $\tilde t_{\rm peak} = 0.87$. The gray bands show the error in prediction propagated from the error in estimation of $\tilde t_{\rm peak}$.

In the model-agnostic approach the regression is completely data driven, and hence, does not represent any underlying principle of the disease spread. This can be considered as a drawback of this approach. However, we show that the predictions made through this approach is very similar to the one made with the BSM. In addition the fit to data is also quite robust as can be seen from \autoref{fig:pred}. 

\subsection*{Discussions on the predictions}

There are several subtleties about the prediction that we should mention here. Firstly, the most important and sensitive part of the prediction is locating the peak of $dN(t)/dt$ during an ongoing phase. For some cases $t_{\rm peak}$ is easy to locate, for instance, for Italy and Ireland in \autoref{fig:pred}. However, for some other countries like Germany, it is somewhat more involved. Secondly, just past the peak it is difficult to gauge whether $N(t)$ will be monotonically decreasing or not. Lastly, in cases like Germany and UK we expect a significant margin of error in determining the peak itself since the $dN(t)/dt$ is not strictly unimodal. This uncertainty in the determination of the peak is what the error bars for the predictions quantify. The error in determining $t_{\rm peak}$ propagates to an error in determining $N_{\rm max}$. Hence, the error bars are small for Ireland and Italy, while they are quite large for Germany and California---the latter being just past the peak in the dataset we use. The spread of COVID-19 in Russia had just reached the peak while we were doing this analysis and hence, we make pure predictions without any test data.

In all the plots in \autoref{fig:pred}, we see that Type II transmission lasts much longer than Type I transmission. For Ireland we see that it was proceeding along the curve for Type I transmission until the next phase started when it diverged once more. Form this plot we now get an idea of how the pandemic might have been routed in Ireland had the next wave not started. UK on the other hand seemed to be following the trajectory taken by Type II transmission until the number of new cases per day started rising once more.

\section*{Summary}
\label{sec:summary}

The surfacing of a worldwide phenomenon such as a pandemic is highly disruptive. The social, political, and economic disruption that this pandemic has brought resulted in an essentially chaotic response to its mitigation. Everything from testing strategies to containment procedures followed very different patterns in different nations and variation even at the regional levels were very palpable. This has made the analysis of the data to model the underlying dynamics of disease spread a formidable task that has rarely yielded to any generalization.

In this paper we show that the emergence of order from within this chaos is made possible by applying well tested procedures drawn from the study of critical phenomenon leading to the formation of universality classes. Such ordering in the data leads to the possibility of characterizing the transmission dynamics into two classes, distinguished by how well the disease has been contained by each country or region. This clearly shows that the variation of the transmission dynamics, and the variation in policies related to disease mitigation, testing guidelines, exit strategies can be scaled away to instigate a data collapse into two classes that allows us to bring about remarkable predictability in the spread of the disease.

We reemphasize that without having to go into the rigours of extracting epidemiological parameters and having to formulate fine-tuned models for the spread of the disease, the emergence of universality allows for the prediction of future disease spread. 

The lack of a definitive understanding of what model to use and how to extract epidemiological parameters has been a major hurdle to making mathematically sound predictions during the spread of COVID-19. In addition, such methods require that each country and each region be treated separately, quickly leading to a proliferation of parameters and systems of equations. While being model-agnostic certainly has several drawbacks, the universality in the spread of the disease that we bring to surface here circumvents many of these problems by exploring the emergence of one single functional form that can be used to make prediction for the spread of the disease once a region has reached the peak of the spread. This clearly allows policy decisions to be made around exit strategies and lets the responsible governance understand how long it might take for a spreading wave of the disease to be contained given the current measures are kept in place.

We would like to point out that the advent of vaccination can be accommodated for in our analysis since it effectively reduces the size of the total susceptible population which, in turn, does not affect the rescaled parameters since they are scaled by $N_{\rm max}$. Therefore, the validity of our work is not short-lived and have general applicability, regardless of the mitigation methods used.

Lastly, we bring to light the proposition of the BSM which explains the data remarkably well. We wish to highlight that it is an intriguing coincidence that Type II dynamics should live at almost the edge of blue sky bifurcation in the BSM. Looking at the finite nonzero value of the order parameter vis-\`a-vis the almost vanishing value of order parameter for Type I transmission, one is readily reminded of the phase transition phenomenon in physical systems. It is tempting to interpret the two types of transmission as two phases quantified by an order parameter. Furthermore, it remains to be seen if the newly introduced BSM finds applications in other epidemiological systems.

\section*{Methods and Analysis Procedures}

All the numerical analysis in this work was performed using Python 3 and several libraries like SciPy, NumPy, Pandas and statsmodels.  All the code necessary for reproducing our results, further exploring the data, making predictions and for making the plots are available in a GitHub repository: \href{https://github.com/talismanbrandi/Universality-COVID-19}{https://github.com/talismanbrandi/Universality-COVID-19}. 

\subsection*{Data sources and curation}

In this work we used data from publicly available sources. The two primary sources that we used are:

\begin{itemize}
    \item Data on COVID-19 confirmed case rate obtained from the Johns Hopkins University, Center for Systems Science and Engineering database~\cite{Dong:2020ka} through their \href{https://github.com/CSSEGISandData/COVID-19}{GitHub repository}. The time-series for all the countries in the world and for all the counties in the USA are available from here. For the confirmed case rates in each state of the USA we merged the data for the counties in each state.
    \item For Germany the data is available through the Robert Koch Institute from the \href{https://npgeo-corona-npgeo-de.hub.arcgis.com/datasets/dd4580c810204019a7b8eb3e0b329dd6_0/data}{RKI COVID19 database}.
\end{itemize}

For all datasets we worked with the seven day averages to remove the fluctuations due to variations in reporting over the period of a week. The last date on the time-series used in this work is 31$^{st}$ of December 2020.

\subsection*{Rescaling of the BSM Differential Equations}

The rescaled BSM can be explicitly written as:
\begin{align}
    \frac{d\tilde N(\tilde t)}{d\tilde t} &= \tilde\beta \tilde N(\tilde t)\left(1 - \frac{\tilde N(\tilde t)}{\tilde\kappa_{\rm lgm}} + \tilde\alpha \tilde N(\tilde t)^2\right).\label{eq:bsma}
\end{align}
For the purpose of prediction, we need to estimate $\tilde t = \tilde t_{\rm peak}$ at the peak of the distribution of the first derivative, $d\tilde N/d\tilde t$. For Type I transmission, $\tilde t_{\rm peak} = 1$ to a very good approximation. Numerically, $\tilde t_{\rm peak} = 0.98$ from the fit shown in \autoref{fig:universal} for Type I transmission since the solution lies very close to the pure logistic growth model ($\tilde\alpha=0$). However, for Type II transmission, $\tilde t_{\rm peak}$ depends on the fit parameters. Since the transmission happens at the bifurcation point
---$\tilde \alpha = 1/4\tilde\kappa_{\rm lgm}^2$---integration of \autoref{eq:bsma} at the corresponding value of parameters is needed. Integration yields:
\begin{align}
    \tilde t = \frac{1}{\tilde\beta}\left[\ln\left|\frac{\tilde N(\tilde t)(\tilde N_0-2\tilde\kappa_{\rm lgm})}{\tilde N_0(\tilde N(\tilde t)-2\tilde\kappa_{\rm lgm})}\right| - \left(\frac{2\tilde \kappa_{\rm lgm}}{\tilde N(\tilde t)-2\tilde \kappa_{\rm lgm}} - \frac{2\tilde \kappa_{\rm lgm}}{\tilde N_0-2\tilde\kappa_{\rm lgm}}\right)\right].
    \label{eq:BSMt}
\end{align}
where $\tilde N_0 = \tilde N(\tilde t=0)$. 

In general, the BSM has three extrema in $d\tilde N/d\tilde t$ at $\tilde N(\tilde t) = 0$, $\tilde N(\tilde t) = 2\tilde \kappa_{\rm lgm}/3$, and $\tilde N(\tilde t) = 2\tilde\kappa_{\rm lgm}$. For the solution at the bifurcation point, there are only two extrema which are the first two, and the maximum is then at $\tilde N(\tilde t) = 2\tilde \kappa_{\rm lgm}/3$. Putting all of these together, one finds from \autoref{eq:BSMt},
\begin{align}
    \tilde t_{\rm peak} = \frac{1}{\tilde\beta}\left[\ln\left(\frac{2\tilde\kappa_{\rm lgm}-\tilde N_0}{2\tilde N_0}\right) + \frac{2\tilde\kappa_{\rm lgm} - 3\tilde N_0}{4\tilde\kappa_{\rm lgm} - 2\tilde N_0}\right].
    \label{eq:BSMtmax}
\end{align}
On using the fit parameters for transmission of Type II shown in \autoref{fig:universal}, we get $\tilde t_{\rm peak} = 0.85$ from \autoref{eq:BSMtmax}. Since $\tilde N_0 \ll \tilde\kappa_{\rm lgm}$, we can simplify the expression for $\tilde t_{\rm peak}$ to:
\begin{align}
    \tilde t_{\rm peak} = \frac{1}{\tilde\beta}\left[\ln\left(\frac{\tilde\kappa_{\rm lgm}}{\tilde N_0}\right) + \frac{1}{2}\right].
    \label{eq:BSMtmaxsim}
\end{align}

\subsection*{Numerical solution of BSM}

The fit to the BSM was performed by numerical integration of the solution to the differential equation given by \autoref{eq:bsma}. For this we used the SciPy package~\cite{2020SciPy-NMeth}. 

\subsection*{Deep neural network architecture}

To construct the DNN used in the model-agnostic fit we used TensorFlow~\cite{tensorflow2015-whitepaper}. The following are the details of the DNN architecture:

\begin{itemize}
\itemsep0em
    \item depth: 3 layers
    \item width: 16 nodes in each layer
    \item optimizer: Adam with a learning rate of 0.01
    \item loss function: mean square error
    \item regularization: early stopping
    \item validation split: 0.2
\end{itemize}

Since the data for $\tilde N (\tilde t)$ is already scaled between 0 and 1, no scaling of the data was necessary. The DNN took about 100-200 epochs to train with the regression taking only a few seconds to be completed. A more detailed discussion of the architecture can be found in the following sections.

\subsection*{Algorithm for predictions}
\label{sec:algopred}

Once we have the fit to the BSM or have trained the DNN with phase-1 data we can make predictions for the subsequent phases. The procedure we follow for this is given by:

\begin{enumerate}
\itemsep0em
    \item Choose a starting date where the cases start to rise (the start-date) from the plot of $dN(t)/dt$ for the second or any subsequent phase.
    \item Subtract the number of cases up until the start-date from the total number of cases to get the adjusted number of cases.
    \item Locate the time $t = t_{\rm peak}$ at which the peak of  $dN(t)/dt$ occurs with $t=0$ at the start-date.
    \item The time $t_{\rm peak}$ is proportional to $t_{1/2}$ with the constant of proportionality being approximately 1 for Type I transmission and a constant (approximately 0.85) extracted from data for phase 1 of the spread for transmission of Type II.
    \item Find the adjusted number of cases on the date corresponding to $t_{1/2}$. Twice that number is the adjusted $N_{\rm max}$.
    \item Use the functions fit from data for phase 1 (the BSM for model-dependent predictions and DNN for model-agnostic predictions) to estimate the future adjusted number of cases.
    \item Now add back the number of cases up until the start-date to get the prediction of the disease spread.
\end{enumerate}

\section*{Acknowledgements}
This research was supported in part through the Maxwell computational resources operated at DESY, Hamburg, The work of A.~Paul is partly funded by Volkswagen Stiftung, Germany, grant no. 99091 for the ``Talisman'' project. We would like to thank Jon Crowcroft, Alejandro de la Puente, Mikhail Prokopenko and Melinda Varga for useful comments.

\section*{Appendix: Universality in physical systems}
\label{sec:emergence}

\begin{figure*}[ht!]
	\centering
	\includegraphics[width=0.49\textwidth]{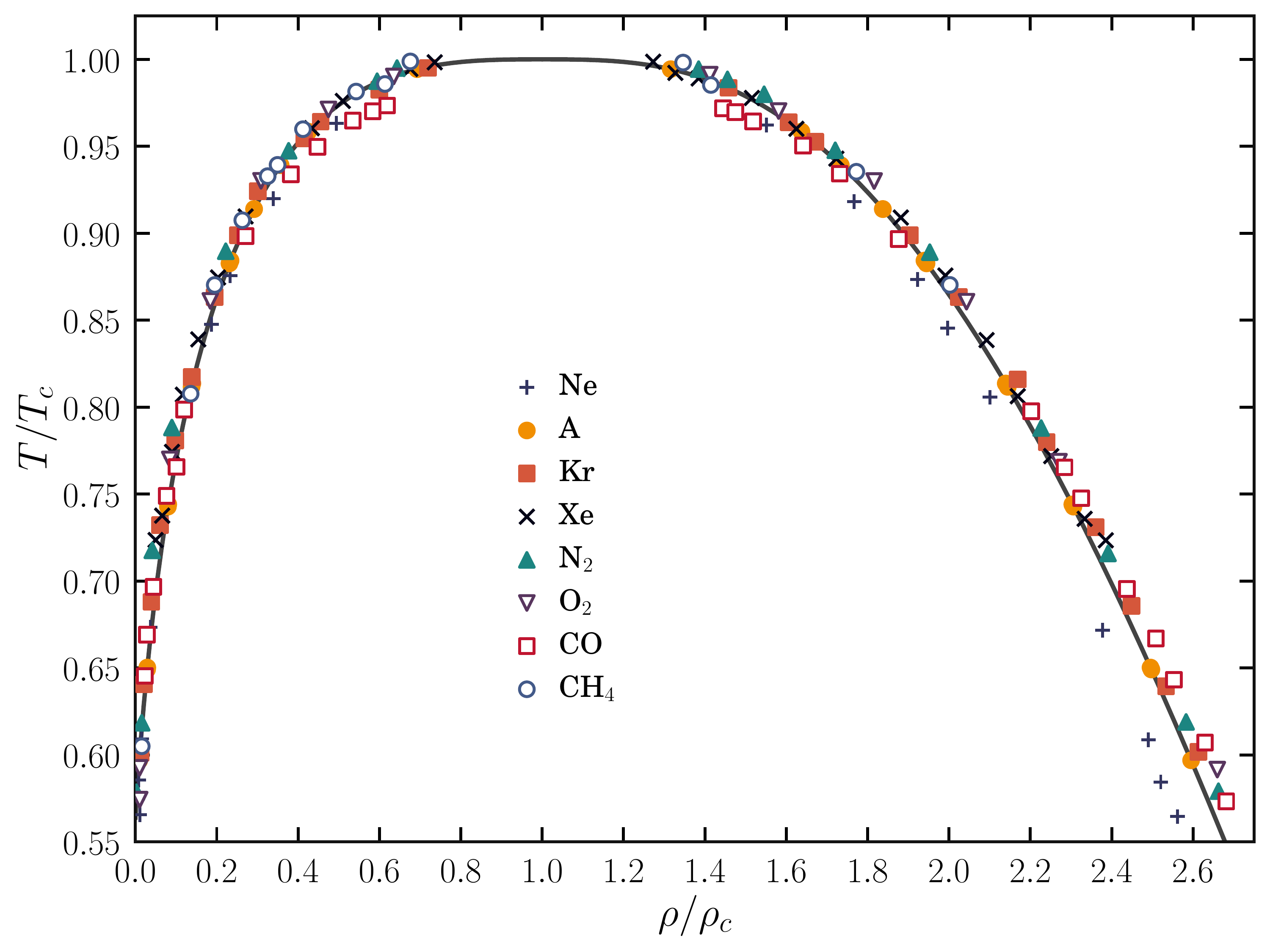}
	\includegraphics[width=0.49\textwidth]{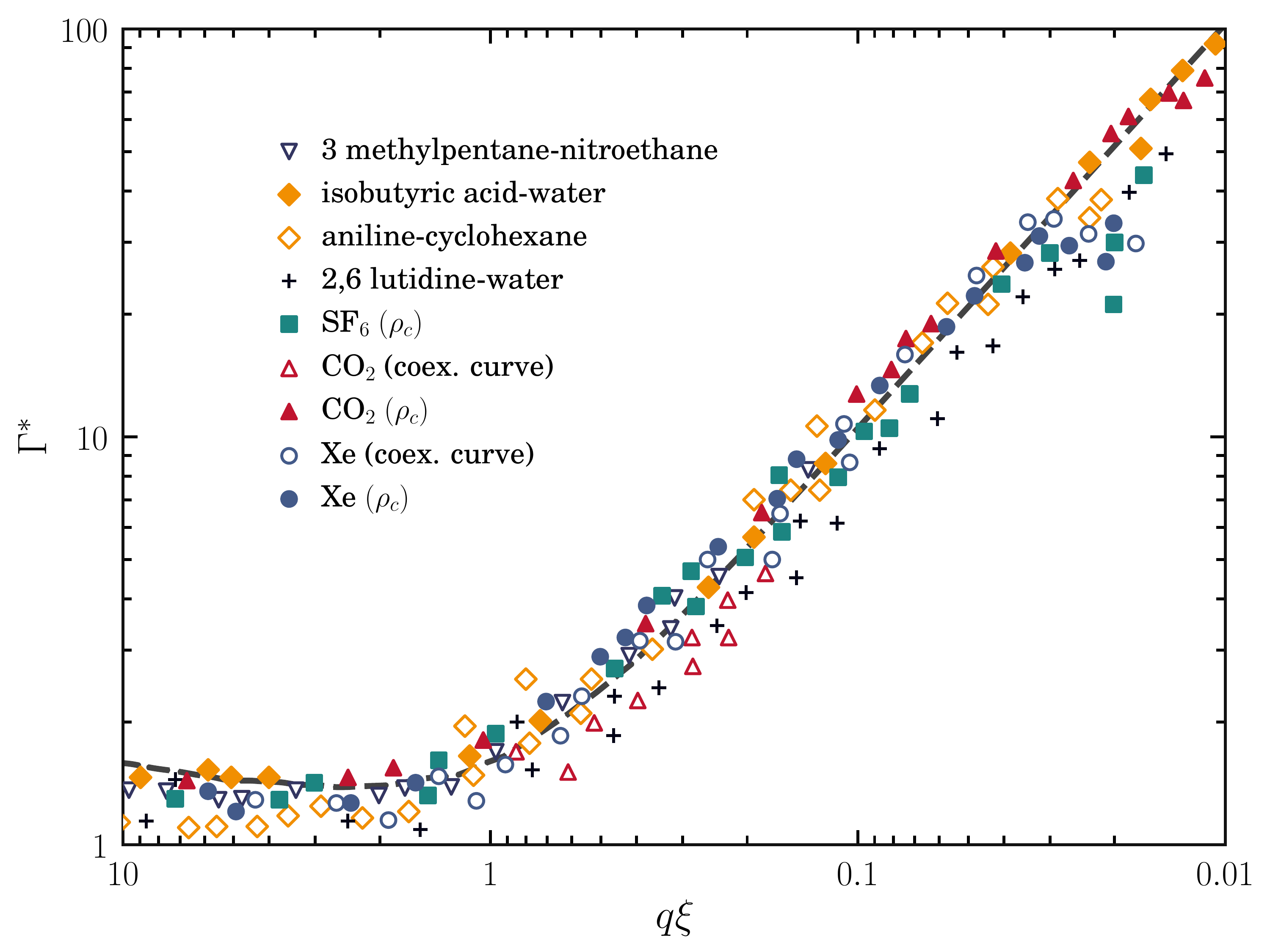}
	\caption{\it \textbf{The emergence of universality in fluid systems}.The {left panel} is a reproduction of figure 2 of Ref.~\cite{doi:10.1063/1.1724033} which shows data collapse in the coexistence of liquid and vapour states for several fluids. The {right panel} is a reproduction of figure 10 of Ref.~\cite{PhysRevA.8.2586} and shows dynamical scaling for several fluids.}
	\label{fig:scaling}
\end{figure*}

In this section we discuss some examples of data collapse seen in physical system. The emergence of universality accompanied by a data collapse is quite common and has been observed for more than half a century. We begin our exposition with data collapse in the study of critical phenomenon~\cite{RevModPhys.49.435,RevModPhys.71.S358,doi:10.1142/S0217979204027748} in fluids. We intend to keep the discussion only at a representative level to give some introduction to universality and data collapse and not make it an exhaustive survey which is not within the scope of this paper.

It is well known that if a gas is subjected to isothermal compression, then at a certain pressure it undergoes a phase transition to the liquid state while the pressure remains constant. Upon completion of the phase transition, the pressure increases very sharply on further compression while the volume increases too. If the temperature is raised and the process repeated, the same phenomenon occurs except that the parameter space over which the pressure and volume simultaneously rise grows smaller and smaller as the temperature is raised. At a particular temperature, $T=T_c$, this parameter space reduces to a point and at higher temperatures the liquefaction does not occur however high the pressure is. The temperature, $T_c$, is the critical temperature and in its vicinity the fluid shows a large amount of fluctuation as it lies between a state of reasonable order, the liquid state, and a state of disorder, which is the gaseous state.

The fluctuations in density very close to the critical point actually mean that some short-lived liquid state bubbles are  produced in the gas. As the critical point is reached these fluctuations become long lived and infinitely long ranged as one has a liquid-vapour coexistence. The infinite lifetime of fluctuations and the infinite range of correlation mean that this phenomenon cannot be sensitive to which gas is being considered. Whether it be Carbon Dioxide, Oxygen, Nitrogen, Xenon or Sulphur Hexafluoride, they will all behave in the same way. This is known as universality and, hence, means that if we draw the equation of state of any gas near the critical point with pressure scaled by the critical pressure and the volume scaled by the critical volume, then we will get just one curve. The critical volumes, critical temperatures and critical pressures are all different for the different materials but the scaled equation of state is the same as can be seen from \autoref{fig:scaling}. If $P$, $V$ and $T$ are the pressure, volume and temperature of a fluid with $a$ and $b$ being constants specific to the fluid, then
\begin{equation}
\left(p+\frac{a}{V^2}\right)\left(V-b\right) = RT \longrightarrow \left(\pi+\frac{3}{\phi^2}\right)\left(3\phi-1\right) = 8\tau,
\label{eq:vanderwallscaling}
\end{equation}
where $\pi=P/P_c$, $\phi=V/V_c$ and $\tau=T/T_c$ with $P_c$, $V_c$ and $T_c$ being the pressure volume and temperature at the critical point. The equation of state thus collapses to a single equation independent of the nature of the fluid.

An identical situation holds for the dynamics which probes the lifetime of the fluctuations. The experimental technique for probing fluctuations directly is frequency resolved light scattering. The intensity $I$ as a function of frequency, $\omega$, is what is measured and at a given temperature and has a Lorentzian shape. As the temperature is lowered towards the critical point, both the peak (at $\omega=0$)  and the width at half-maximum increase with decreasing temperature, both headed for infinitely large values at the critical point. Different materials show different Lorentz distributions at different temperatures. If we denote the intensity at a particular frequency and a particular temperature, $T$, by $I(\omega,\Delta T)$, with $\Delta T=T-T_c$, then for different materials and  different sets of $\Delta T$ one has different $I(\omega,\Delta T)$ vs $\omega$ distributions. However, if one plots $I(\omega,\Delta T)/I(0,\Delta T)$  vs. $\omega/\omega_0$  where $\omega_0$ is the characteristic frequency for a given material, then the data for hundreds of individual distributions collapse on a single distribution. This phenomenon is called dynamic scaling~\cite{RevModPhys.49.435,RevModPhys.71.S358,doi:10.1142/S0217979204027748} and is displayed in \autoref{fig:scaling}.

This phenomenon of universal data collapse is ubiquitous: It appears in various forms in network dynamics~\cite{BarzelNP}, neuronal network~\cite{PhysRevLett.108.208102}, brain~\cite{10.2307/3067747, Serafinoe2013825118}, supercritical fluids~\cite{JPCL}, amorphous solids~\cite{Ulrich20929}, Bose--Einstein condensate~\cite{Klinder3290}, jamming transition~\cite{Goodrich9745}, granular media~\cite{Manna_1991,Denisov}, financial markets~\cite{PhysRevE.94.042305},  metallic liquids~\cite{Blodgett}, etc. Thus, in some sense, it is not surprising that this intriguing phenomenon be present in the transmission of infectious diseases as well.

\section*{Appendix: Comparisons between different models}
\begin{figure*}[ht!]
    \centering
    \includegraphics[width=0.9\textwidth]{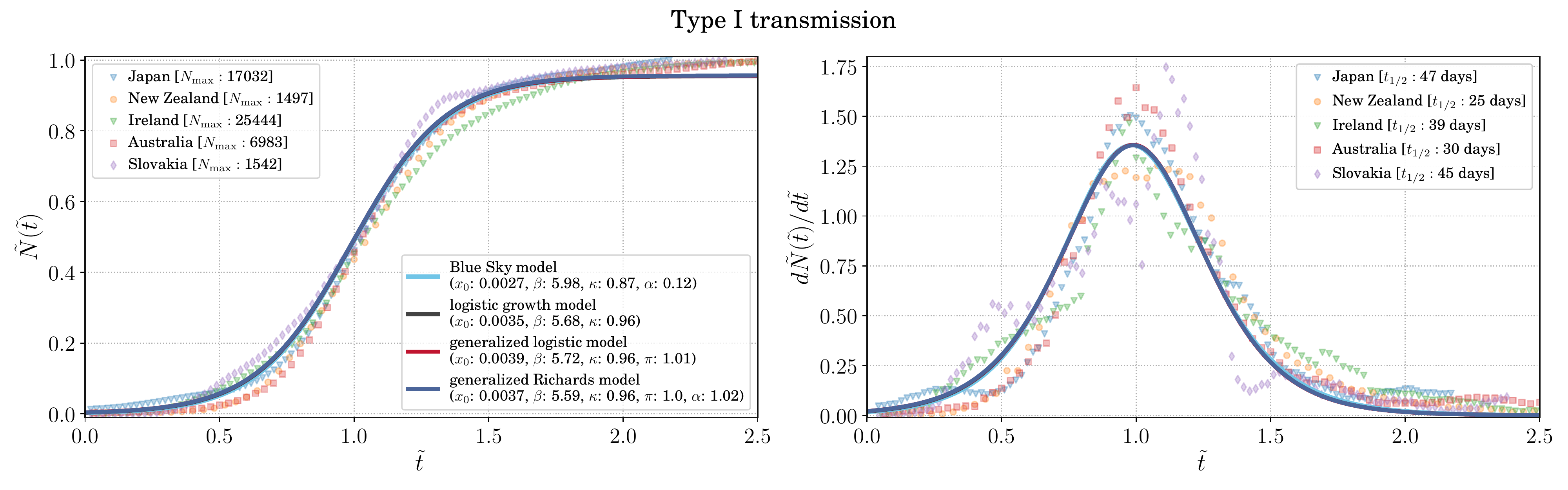}
    \includegraphics[width=0.9\textwidth]{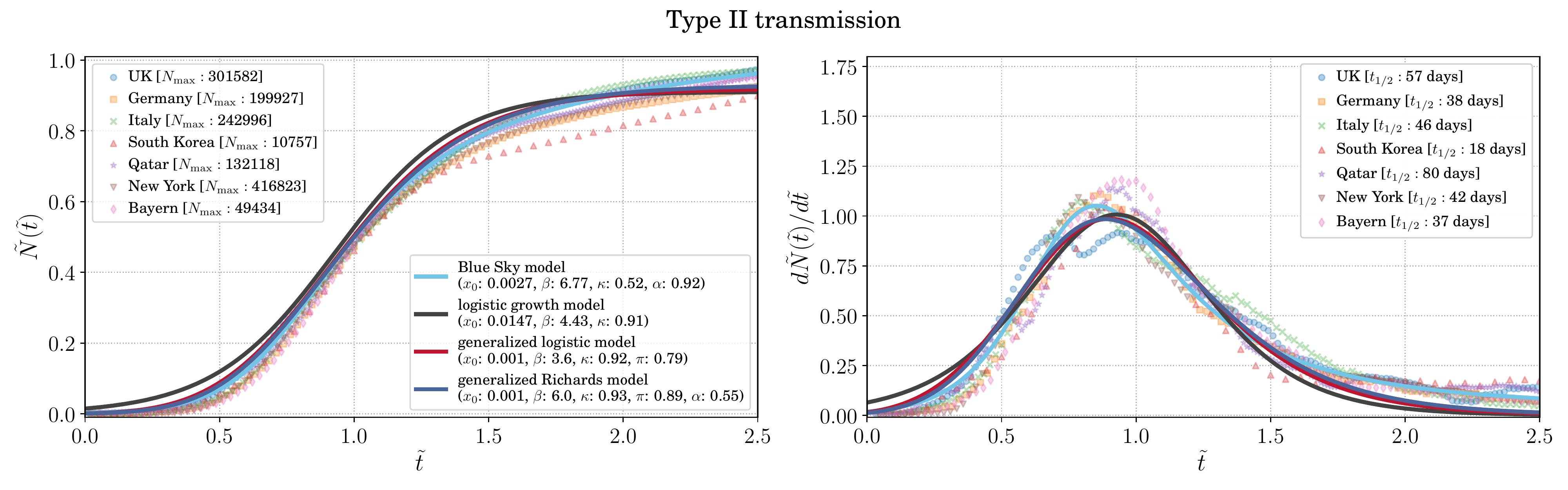}
    \caption{\it \textbf{A comparison of the variants of the logistic model.} The panels on the left show  $\tilde N(\tilde t)$ vs. $\tilde t$ distribution for Type I and Type II transmission dynamics; and the right panels shows the $d\tilde N(\tilde t)/d\tilde t$ vs. $\tilde t$ plot for the same. We fit four models to the data -- the logistic growth model, the generalized logistic model, the generalized Richards model and the Blue Sky model. While the fit to data for all the models are the same for Type I transmission where all the models go to the limit of being a logistic growth model, for Type II transmission the Blue Sky model can explain the data better than the other models.}
    \label{fig:SI_allmodel}
\end{figure*}

In this section we compare some models that have been used commonly in epidemiology to quantify transmission dynamics of diseases including COVID-19~\cite{cite-key3}. All the models that we describe here are variants of the logistic growth model. The logistic growth model (LGM) is defined as:
\begin{align}
    \frac{dN(t)}{dt} &= \beta N(t)\left(1 - \frac{N(t)}{\kappa} \right).
\end{align}
We define $\tilde t = t/t_{1/2}$ and $\tilde N (t) = N(t)/N_{\rm max}$, and by setting $\tilde\beta = \beta t_{1/2}$ and $\tilde\kappa = \kappa/N_{\rm max}$, we get the rescaled equation:
\begin{align}
    \frac{d\tilde N(\tilde t)}{d\tilde t} &= \tilde\beta \tilde N(\tilde t)\left(1 - \frac{\tilde N(\tilde t)}{\tilde\kappa} \right).
\end{align}
We see that $\beta$ scales as the time dimension and $\kappa$ scales as the number of cases. This makes the dimensional quantities, $\tilde\beta$ and $\tilde\kappa$, independent of any characteristic timescale or size of the system. $\tilde\kappa = 1$ in the LGM.

The generalized logistic model (GLM) is given by:
\begin{align}
    \frac{dN(t)}{dt} &= \beta N(t)^\pi\left(1 - \frac{N(t)}{\kappa} \right).
\end{align}
Similar rescaling as above leads to the rescaled parameters $\tilde\beta = \beta t_{1/2}N_{\rm  max}^{\pi -1}$ and $\tilde\kappa = \kappa/N_{\rm max}$ leading to the rescaled GLM being:
\begin{align}
    \frac{d\tilde N(\tilde t)}{d\tilde t} &= \tilde\beta \tilde N(\tilde t)^\pi\left(1 - \frac{\tilde N(\tilde t)}{\tilde\kappa} \right).
\end{align}

The generalized Richards model (GRM) is given by:
\begin{align}
    \frac{dN(t)}{dt} &= \beta N(t)^\pi\left[1 - \left(\frac{N(t)}{\kappa}^\alpha\right) \right].
\end{align}
For $\pi=1$, it reduces to the standard Richards model~\cite{10.1093/jxb/10.2.290}. We see that the scaling is similar to the GLM with $\tilde\beta = \beta t_{1/2}N_{\rm  max}^{\pi-1}$ and $\tilde\kappa = \kappa/N_{\rm max}$; the rescaled GRM is:
\begin{align}
    \frac{d\tilde N(\tilde t)}{d\tilde t} &= \tilde\beta \tilde N(\tilde t)^\pi\left[1 - \left(\frac{\tilde N(\tilde t)}{\tilde\kappa}\right)^\alpha \right].
\end{align}

In \autoref{fig:SI_allmodel} we show a comparison between all the models we list here and the Blue Sky model (BSM). The two upper panels are for transmission of Type I. The LGM fits this data quite well and all the other models acquire the parameter values that take them to the limit of being an LGM.  The two lower panels are for transmission of Type II. Here we clearly see that the BSM fits the data much better that the other models. Especially for the tail of distribution past the peak of the $d\tilde N(\tilde t)/d\tilde t$ vs. $\tilde t$ plot where $\tilde t > \tilde t_{\rm peak}$, the BSM model allows for larger $d\tilde N(\tilde t)/d\tilde t$ compared to the other models hence being able to replicate transmission of Type II where the spread of the disease lingers on.

\section*{Appendix: The deep neural network architecture}
\begin{figure}[h!]
    \centering
    \includegraphics[width=0.6\textwidth]{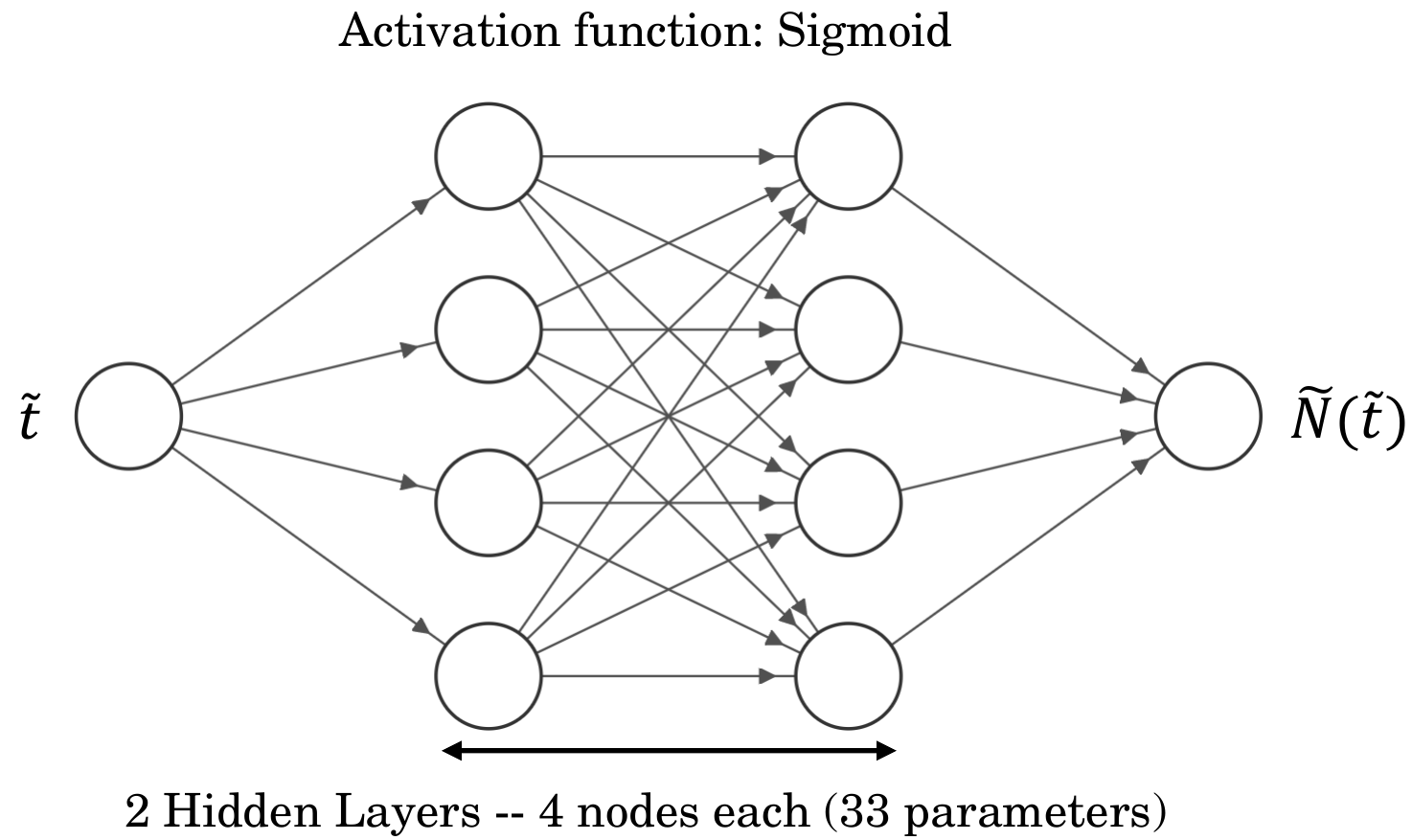}
    \caption{\it \textbf{A representative architecture of the DNN used for the model-agnostic analysis to fit $\tilde N(\tilde t)$ as a function of $\tilde t$}. For the final analysis we used a similar architecture with 3 layers each with 16 nodes to keep the analysis as model-agnostic as possible. To regulate the training of the DNN we used an early callback procedure which stops the training when the validation loss stops decreasing thus avoiding training biases.}
    \label{fig:dnn}
\end{figure}

In our model-agnostic approach we used a deep neural network (DNN)~\cite{6472238,SCHMIDHUBER201585} to fit the data. A DNN is a universal function generator. The reason we chose a DNN over other machine learning frameworks (i.e. decisions trees, random forests, SVM etc.) is the ease with which it can be use for a regression to noisy data to extract the underlying function. We will describe in brief the choice of the network architecture.

We started with a small DNN with 1 input node for $\tilde t$, 2 hidden layers, each with 4 nodes and 1 output layer for $\tilde N (\tilde t)$. All layers were fully connected as shown in \autoref{fig:dnn}. The activation function used was the $sigmoid$ function. The solution to the LM is a $sigmoid$ function and hence the transmission of Type 1 can be fairly explained with a just one sigmoid function that connects the input $\tilde t$ to the output $\tilde N (\tilde t)$. This is not possible though for transmission of Type II which shows clear deviation from the LM. However, in the model agnostic approach we did not want to introduce any model biases through assumptions of functional form. Hence we used a larger (and deeper) neural network with 33 parameters to begin with. We scaled up the network to 3 hidden layers each with 16 nodes to check for the stability of our approach. The regression with the DNN with 2 hidden layers with 4 nodes each gave almost the same fit as the one with 3 hidden layers with 16 nodes each having a total of 593 parameters. To keep our method as general as possible and not introduce model biases from the activation function we decided to use the larger DNN with 3 hidden layers each having 16 nodes.

In addition, we tested the DNN with other activation function like $tanh$ and $relu$. For $tanh$ a somewhat wider and deeper network was needed to fit to the data. With the $relu$ activation the network needed to be very large. Hence, we used the $sigmoid$ activation function for the final DNN architecture.

Even though the DNN we used was not so large we used an early callback algorithm to stop the training of the network when the validation loss stopped decreasing to avoid overtraining and chose the epoch where the validation loss was the least. The loss function that we used was a simple mean square error function which measures the $L_2$ distance of the fit from the data. The training time was a few seconds with the early stopping coming into effect at about 100-150 epochs. The trained DNN was used for making prediction for the ongoing phases

\bibliography{bibliography}

\section*{Author contributions statement}
A. Paul, J.K.B., and S.C. designed and performed research; and wrote the paper. A. Paul wrote all the codes necessary for processing and analyzing the data. A. Pal took part in preliminary analytical investigations of the data and models.


\section*{Data and materials availability}
All data and materials necessary to reproduce this work is available in a GitHub repository:\newline \href{https://github.com/talismanbrandi/Universality-COVID-19}{https://github.com/talismanbrandi/Universality-COVID-19}.

\end{document}